\documentclass[prl,amsmath,amssymb,superscirptaddress,reprint,floatfix,longbibliography]{revtex4-1}
\usepackage{graphicx}% Include figure files
\usepackage{bm}
\usepackage{amsmath,amssymb}
\usepackage{color}
\usepackage[caption=false]{subfig}
\usepackage{enumitem}
\usepackage{lipsum}

\newcommand{\ket}[1]{\ensuremath{|{#1}\rangle}}

\newcommand{\maxenh}{\beta_0}
\newcommand{\enh}{\beta}
\newcommand{\Dc}{\Delta}
\newcommand{\Omc}{\Omega}
\newcommand{\Dr}{\Delta_\mathrm{r}}
\newcommand{\Omr}{\Omega_\mathrm{r}}
\newcommand{\sigr}{\sigma_\mathrm{r}}
\newcommand{\delr}{\delta_\mathrm{r}}
\newcommand{\gamr}{\gamma_\mathrm{r}}
\newcommand{\gams}{\gamma_\mathrm{sg}}
\newcommand{\Gc}{\Gamma}
\newcommand{\Gr}{\Gamma_\mathrm{r}}
\newcommand{\kr}{k_\mathrm{r}}
\newcommand{\vT}{v_\mathrm{T}}
\newcommand{\ODb}{\mathrm{OD}_{\mathrm{B}}}

\begin{document}

\title{Recovering the homogeneous absorption of inhomogeneous media}

\author{Ohr Lahad}
\thanks{These authors contributed equally to this work.}
\affiliation{Department of Physics of Complex Systems, Weizmann Institute of Science, Rehovot 76100,
Israel}
\author{Ran Finkelstein}
\thanks{These authors contributed equally to this work.}
\affiliation{Department of Physics of Complex Systems, Weizmann Institute of Science, Rehovot 76100, Israel}
\author{Omri Davidson}
\affiliation{Department of Physics of Complex Systems, Weizmann Institute of Science, Rehovot 76100, Israel}
\author{Ohad Michel}
\affiliation{Department of Physics of Complex Systems, Weizmann Institute of Science, Rehovot 76100, Israel}
\author{Eilon Poem}
\affiliation{Department of Physics of Complex Systems, Weizmann Institute of Science, Rehovot 76100, Israel}
\author{Ofer Firstenberg}
\affiliation{Department of Physics of Complex Systems, Weizmann Institute of Science, Rehovot 76100, Israel}

\begin{abstract}
\noindent
The resonant absorption of light by an ensemble of absorbers decreases when the resonance is inhomogeneously broadened. Recovering the lost absorption cross-section is of great importance for various applications of light-matter interactions, particularly in quantum optics, but no recovery mechanism has yet been identified and successfully demonstrated. Here, we formulate the limit set by the inhomogeneity on the absorption, and present a mechanism able to circumvent this limit and fully recover the homogeneous absorption of the ensemble. We experimentally study this mechanism using two different level schemes in
atomic vapors and demonstrate up to 5-fold enhancement of the absorption above the inhomogeneous limit.
Our scheme relies on light shifts induced by auxiliary fields and is thus applicable to various physical systems and inhomogeneity mechanisms.

\end{abstract}

\maketitle

Inhomogeneous broadening of spectral lines is a prevalent limiting factor in experiments and applications involving light-matter interactions in ensembles. This common impediment occurs for various atomic and atom-like absorbers, including quantum dots \cite{Woggon2002QDs, Steel2005QDsRaman, Alivisatos1999QDs}, diamond color-centers \cite{atature2014single,jelezko2014single}, rare-earth ions in crystals \cite{Gisin2010telecom,Qreview2010}, hot atoms \cite{Kitching2005}, and particularly with Rydberg excitations \cite{carr2012three,Pfau2018SinglePhoton}. The broadening originates from a distribution of resonant frequencies of the individual absorbers. Common sources of inhomogeneity are nonuniform magnetic fields or local crystal strains, or thermal distributions of atomic velocities.

The inhomogeneous broadening dominates when the width of the resonance-frequency distribution $2\sigma$ is larger than the homogeneous linewidth $2\gamma$. This is accompanied by a decrease in absorption, which scales as $\gamma/\sigma$. Standard techniques for circumventing inhomogeneous broadening rely on selective methods,
such as hole burning and Doppler-free configurations,
which resonantly address only a small fraction of the ensemble \cite{demtroder2015laser}.
Another class of techniques is dynamical decoupling, such as spin echo, which suppresses decoherence while engaging the entire ensemble \cite{Kurnit1964echo,Kraus2006CRIB}. 
Additionally, state population transfer can be made efficient for a large fraction of the ensemble using pulsed light-shifts \cite{Bergmann2000SCRAP,Rebic2009SCRAP}.
However, these techniques do not recover the bare absorption lost due to the inhomogeneity.

It is known that far off-resonance operation is mostly insensitive to inhomogeneous broadening, and so are processes relying on slow light, such as light storage \cite{GorshkovIII,Josh2007,FLAME}.
Scattering effects, however, such as absorption or fluorescence, are directly affected by inhomogeneity. Indeed, the effective absorption cross-section and the optical depth decrease in typical systems by 1-3 orders of magnitude \cite{Woggon2002QDs, Steel2005QDsRaman, Alivisatos1999QDs,atature2014single,jelezko2014single,Gisin2010telecom,Qreview2010,Kitching2005,carr2012three,Pfau2018SinglePhoton}. This is particularly important for extreme nonlinear optics, such as the effective photon-photon interaction mediated by Rydberg atoms, which require high effective optical-depth within an interaction range %$r_{\mathrm{B}}$
of only a few micrometers \cite{gorshkov2011blockade,firstenberg2016review,Pfau2012Guide,Durr2018gate}. As the atomic density is practically limited \cite{Durr2014switch,pfau2014molecular,lukin2015molecule}, the effective cross-section controls the strength of the interaction.

In atomic gases, inhomogeneous broadening results from the Doppler effect for different atomic velocities. It was originally suggested by Cohen-Tannoudji \emph{et al.}
% \cite{CCT1978}
that velocity-dependent light-shifts due to an additional field can be utilized to counteract the broadening due to the Doppler effect. This idea has been analyzed \cite{CCT1978,yavuz2013suppression} and realized \cite{CCT1979,CCT1982} in several configurations. Inhomogeneous light-shifts were also used for generating narrow features with multi-photon excitations of hot Rydberg atoms \cite{carr2012three}, and for increasing the coherence time of laser-trapped atoms \cite{kaplan2002suppression}. 
Several proposals focused on counteracting the decrease in amplitude due to the inhomogeneity   \cite{popov2000enhanced,yavuz2013suppression,CCT1979}.
However, and despite its importance, no significant enhancement of the effective absorption cross-section has been experimentally realized so far.

In this work, we first define the `inhomogeneous limit' of the effective cross-section, for a probe field coupled to inhomogeneously-broadened transitions in either two-level or three-level systems. We find that the same limit applies to both cases, even when a coupling field in the three-level case generates a narrow absorption feature for the probe. The latter occurs, for example, in Raman transitions between two sub-states of a ground level.
% [Fig.~1(a)]
We then show that this limit can be surpassed in a four-level system, by utilizing light-shifts due to an additional `recovery' field. Consequently, the probe field can resonantly interact with the majority of the ensemble, and in the ideal case the absorption is \emph{fully recovered}.
We use a hot atomic medium to study this mechanism and demonstrate a significant increase in the effective cross-section, $\times 5$ higher than the inhomogeneous limit. This is the first demonstration of any absorption effect significantly exceeding the inhomogeneous limit.
Finally, to establish the generality of the recovery mechanism, we implement it in a ladder-type level system involving Rydberg states and demonstrate a similar increase of absorption beyond the inhomogeneous limit.

\emph{The inhomogeneous limit.}---
Consider a two-level system with an optical transition $\ket{g}\leftrightarrow\ket{e}$ of homogeneous (half) linewidth $\gamma$. To introduce inhomogeneity, assume that the resonance frequency of this transition is shifted by a parameter $\delta$ which has a Gaussian distribution with a standard deviation $\sigma\gg\gamma$.
% \footnote{\label{note1} The results of this work are qualitatively valid also for non-Gaussian distributions.}
Consequently, the absorption line broadens, and its amplitude reduced by the factor $\maxenh=\sqrt{(2/\pi)} \sigma/\gamma$. 
A typical example is the $D_1$ transition in hot rubidium atoms, where $\gamma=2.875$ MHz. The Doppler width at $50 ^{\circ}$C is $\sigma= k \vT=220$ MHz, where $k$ is the probe wavevector and $\vT$ the RMS thermal velocity, and therefore the absorption is reduced by a factor of $\maxenh\approx 60$ with respect to that of stationary atoms. In the solid state, an important example are silicon-vacancy (SiV\textsuperscript{-}) centers in diamond at a few K \cite{arend2016,becher2018coherent}. These centers have a homogeneous width of $\gamma_{\mathrm{SiV}}\approx 50$~MHz and an inhomogeneous width of $\sigma_{\mathrm{SiV}}\approx 5$~GHz due to varying local strains in the surrounding lattice, which reduces the absorption by a factor of $\maxenh \approx 80$.

\begin{figure}[t] % [b] is a position specifier, sets to the bottom of the page.
\includegraphics[scale=0.25,trim=0.cm 0.1cm 0.cm   0.3cm,clip=true]{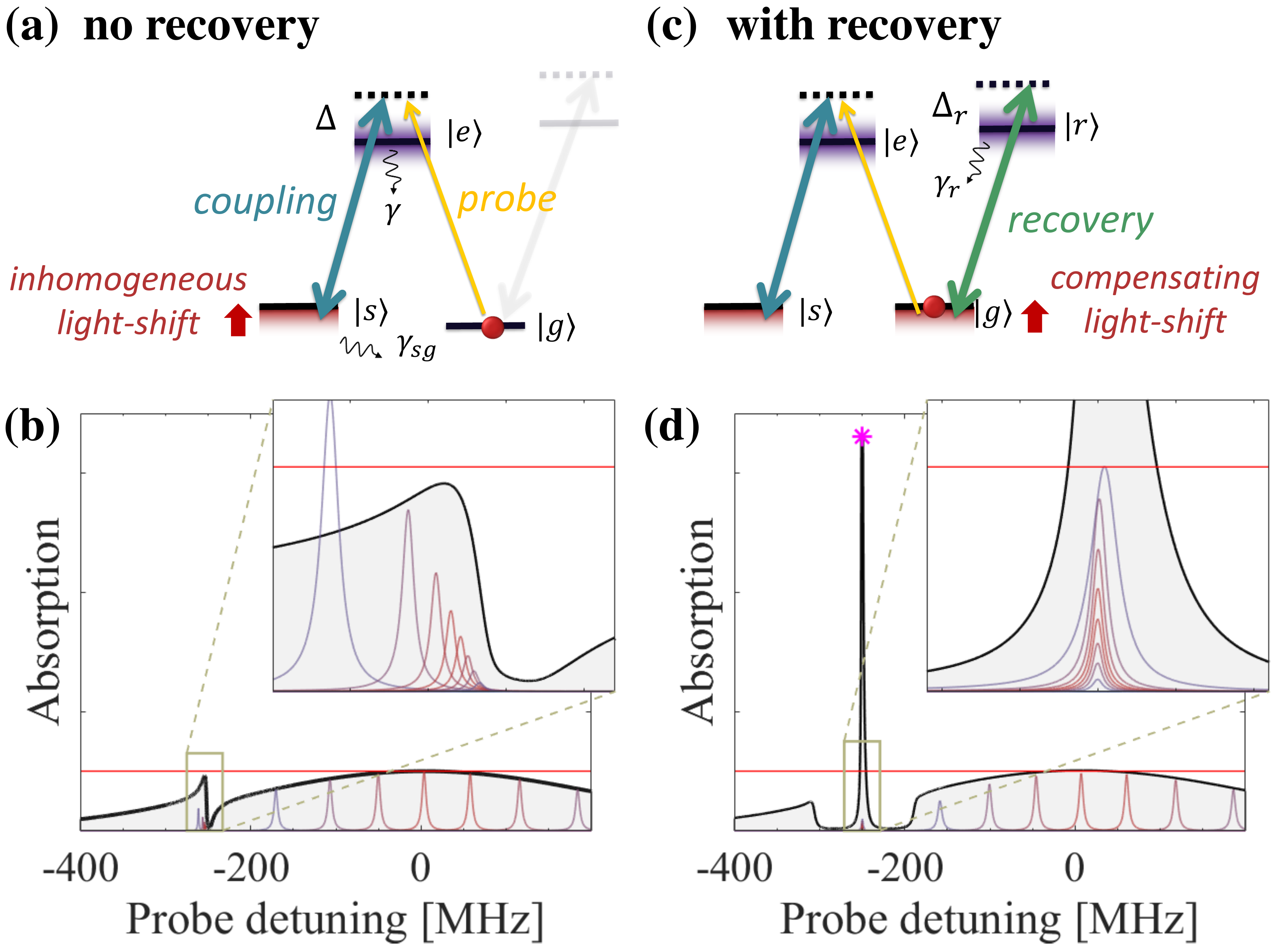}%  trim={<left> <lower> <right> <upper>
	\caption{Induced inhomogeneous broadening (left) and compensation (right). (a) Inhomogeneous broadening of the optical transition $\ket{s}\leftrightarrow\ket{e}$ (purple shading) leads to broadening of the two-photon transition $\ket{g}\leftrightarrow\ket{s}$ (red shading) due to a distribution of light shifts introduced by the far-detuned coupling field. (b) Calculations showing the two-photon resonances for different absorbers (different colors; enlarged in the inset). The resulting absorption spectrum (black), obtained by summing over all absorbers, is bounded by the inhomogeneous limit (red horizontal line). (c) A recovery field driving a second optical transition $\ket{g}\leftrightarrow\ket{r}$ removes the inhomogeneity of the two-photon transition by introducing similar light shifts. (d) With inhomogeneous compensation, the two-photon resonances of different absorbers are aligned (see inset), potentially recovering the absorption of the homogeneous system (magenta asterisk). Along with the merging of the two-photon resonances, a transmission window  akin to the Autler-Townes splitting opens up around the enhanced peak.
	}
	\label{fig_schematic} % label is useful for referring to the figure in the text.
\end{figure}

Now consider a three-level system $\ket{g}-\ket{e}-\ket{s}$, as depicted in Fig.~\ref{fig_schematic}(a), and assume that the frequency difference between $\ket{g}$ and $\ket{s}$ is unaffected by the inhomogeneity. This is common for Raman transitions in $\Lambda$-systems where $\ket{g}$ and $\ket{s}$ are within the ground-level manifold, such that the resonance frequencies of the two optical transitions $\ket{g}\leftrightarrow\ket{e}$ and $\ket{e}\leftrightarrow \ket{s}$ share the same broad distribution of shifts $\delta$. When a coupling field drives the $\ket{e}\leftrightarrow\ket{s}$ transition and is detuned from resonance by $\Dc\gg \sigma$, it can induce the absorption of the probe on the two-photon transition $\ket{g}\leftrightarrow\ket{s}$.
The homogeneous width of the two-photon transition is the sum of its bare width $\gams$ and the scattering rate $(\Omc / \Dc)^2 \gamma$, where $\Omc$ is the Rabi frequency of the coupling field. Notably, both contributions can be substantially smaller than $\sigma$, and thus two-photon transitions are often thought of as circumventing inhomogeneity.

% FIGURE 1

Nevertheless, the absorption of the probe at these two-photon transitions is never higher than that of the broadened one-photon transition. 
For weak coupling fields, this limit is explained by the inefficient coupling $(\Omc / \Dc)^2 \gamma\ll \gams$ to $\ket{e}$. For the less obvious case of strong coupling $(\Omc / \Dc)^2 \gamma\gg \gams$, it results from inhomogeneous  light-shifts:
The strong coupling field shifts the state $\ket{s}$ by $\sim\Omc^2/\Dc$, thereby shifting the two-photon resonance. In an inhomogeneous medium, the light-shift depends on $\delta$ and is given by $\Omc^2/(\Dc-\delta)\approx \Omc^2/\Dc+(\Omc/\Dc)^2\delta.$
It follows that the dressed two-photon transition is inhomogeneously shifted, as shown by the shifted Raman-absorption lines in Fig.~\ref{fig_schematic}(b). Replacing $\delta$ by its standard deviation $\sigma$,
we find an induced inhomogeneous broadening $\sim \left(  \Omc / \Dc \right)^2 \sigma$. The ratio between this broadening and the homogeneous width $\sim \left(  \Omc / \Dc \right)^2 \gamma$ is $\sigma / \gamma$, which is the same ratio as in the two-level case. The amplitude of the two-photon line decreases by the same ratio, and it is therefore bounded by the same inhomogeneous limit. Full density-matrix calculations of an inhomogeneous ensemble confirm that the reduction of absorption is at least $\maxenh$ over the whole spectrum for all choices of $\Omc$ and $\Dc$.

\emph{Inhomogeneous compensation.}---
Our approach for recovering the absorption cross-section is to counteract the induced inhomogeneous broadening by adding another excited state $\ket{r}$ and a \emph{recovery} light-field that drives the optical $\ket{g}\leftrightarrow\ket{r}$ transition [Fig.~\ref{fig_schematic}(c)].
% forming an $N$-type system \cite{Yudin1999TOC,Kim2010EIA,Adams2009Nsystem,Whiting2015EIA}.  
The additional transition is inhomogeneously broadened as well, bearing
frequency shifts $\delr$ that are correlated with $\delta$ and are distributed with standard deviation $\sigr$. The far-detuned recovery field, with detuning $\Dr$ and Rabi frequency $\Omr$, shifts the state $\ket{g}$ by $\Omr^2 / \left( \Dr - \delr \right)$. 
If the light-shifts of $\ket{s}$ and $\ket{g}$
% due to the coupling and recovery fields
are tailored to satisfy the compensation condition
\begin{equation} \label{eqn:compensation_condition}
\frac{\Omc^2}{\Dc-\delta}=\frac{\Omr^2}{\Dr-\delr},
\end{equation}
then the resonance frequency of the two-photon transition will coincide for the entire ensemble, overcoming the inhomogeneity [Fig.~\ref{fig_schematic}(d)].
% We refer to Eq.~\eqref{eqn:compensation_condition} as the compensation condition.

The homogeneous linewidth of the two-photon resonance determines the sensitivity of the recovered absorption to deviations from Eq.~\eqref{eqn:compensation_condition}. This linewidth, averaged over the ensemble, is given by $\Gc+\Gr+\gams$, where the scattering rates due to the coupling and recovery fields are approximately $\Gc=[ \Omc^2/(\Dc^2+\sigma^2)]  \gamma$ and $\Gr=[ \Omr^2/(\Dr^2+\sigr^2)]\gamr$. 

Note that, as long as $\gamma\ll\sigma$, the expressions comprising Eq.~\eqref{eqn:compensation_condition} hold for a large fraction of the ensemble, even when the fields are tuned to within the inhomogeneously-broadened line $|\Dc| \lesssim \sigma$. This near-resonant regime features a wide transmission window surrounding the enhanced peak, whose width of $2\Omc+2\Omr$ is explained by Autler-Townes splittings.

In hot atoms with co-propagating fields, $\delta=k v$ and $\delr=\kr v$ are both linear in the atomic velocity $v$ and therefore correlated; here $k$ and $\kr$ are the wavenumbers of the coupling and recovery fields. The compensation condition is met for $\Omr=\Omc \sqrt{ \kr/k}$ and $\Dr =\Dc \left(\kr/k\right)$.
Other physical systems often exhibit similar correlations. For example, in color-centers such as SiV\textsuperscript{-}, both $\delta$ and $\delr$ depend on the same local crystal strain \cite{atature2014single}.

\emph{Enhancement resource.}---
The compensation condition \eqref{eqn:compensation_condition} guarantees that the resonance frequencies of the two-photon absorption-lines of all the absorbers are aligned. The height of these lines determines the height of the enhanced absorption peak.
We define the enhancement of absorption $\enh$ as the ratio between absorption on the two-photon and one-photon resonances. 
The absorption cross-section can be recovered only up to its homogeneous value, hence the maximal enhancement is $\maxenh$.
Generally, each four-level absorber contributes at its two-photon resonance only a fraction of the homogeneous cross-section. This fraction is determined by the competition between the coupling-field scattering rate $\Gc$ and the total linewidth of the two-photon transition $\Gc+\Gr+\gams$. It follows that the enhancement of absorption is given by $\enh=\maxenh\Gc/(\Gc+\Gr+\gams)$. This intuitive formula agrees extremely well with full density-matrix calculations.

It is instructive to consider the simple case where both excited states ($\ket{e}$,$\ket{r}$) share the same inhomogeneous shifts $\delta=\delr$, and consequently $\sigma=\sigr$, $\Omc=\Omr$, and $\Dc=\Dr$.
% In this case, %one can write
The resulting enhancement is
\begin{equation}\label{eqn:saturation}
  \enh=\maxenh \frac{\gamma}{\gamma+\gamr}\frac{\mu^2}{1+\mu^2}.
\end{equation}
Here $\mu^2=\Omc^2(\gamma + \gamr)/[(\Dc^2+\sigma^2)\gams]$ is a saturation parameter that scales linearly with the coupling and recovery fields intensities, and % we find that
${\enh\to \maxenh\cdot \gamma / \left(\gamma + \gamr   \right)}$ at saturation.
Therefore, full recovery of the homogeneous absorption cross-section requires both a narrow transition for the recovery field $\gamr\ll\gamma$ and a high intensity of the coupling and recovery fields $\mu^2\gg 1$.

% FIGURE 2 % a float is stuck
\begin{figure}[t] % [b] is a position specifier, sets to the bottom of the page.
\includegraphics[scale=0.25,trim=0.cm 6.4cm 0.cm 0.cm,clip=true]{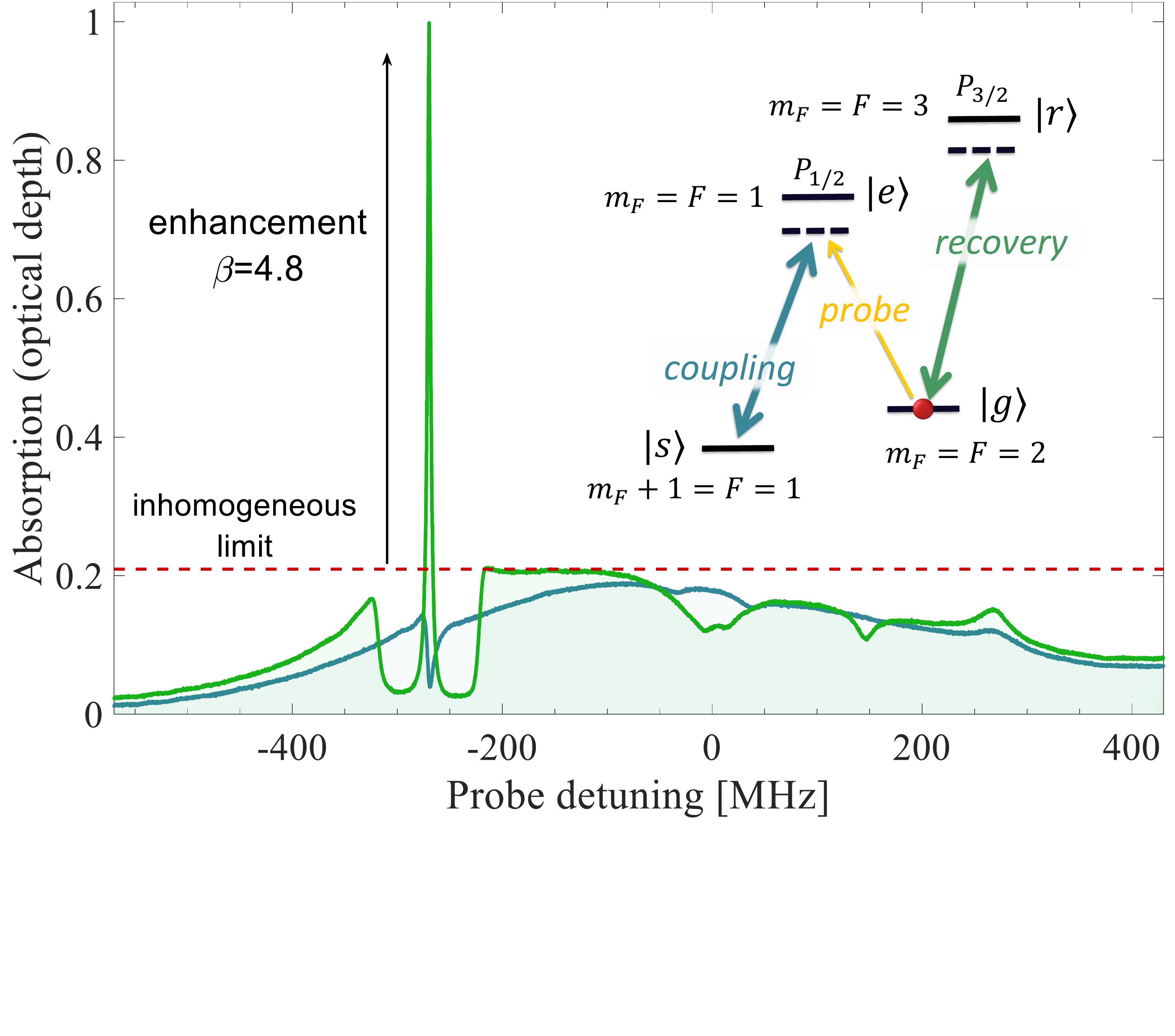} %  trim={<left> <lower> <right> <upper>
	\caption{Experimental demonstration of absorption enhancement with inhomogeneous compensation. The absorption spectra of the probe field exhibit both the one-photon resonance $\ket{g}\to\ket{e}$ near zero detuning, and the two-photon (Raman) resonance $\ket{g}\to\ket{s}$ induced by the coupling field around $\Dc=-270$~MHz. Without the recovery field (blue curve), the absorption on both resonances is bounded by the inhomogeneous limit (dashed red). With the recovery field (green curve), the absorption peak grows substantially, exceeding the inhomogeneous limit by a factor of $\enh=4.8\pm0.4$. In this measurement, $\Omc=29$~MHz, $\Omr=29.6$~MHz, and $\Dr=-300$~MHz.
	}
	\label{fig_spectrum} % label is useful for referring to the figure in the text.
\end{figure}

It is important to distinguish between the mechanism we study and those developed for enhancing the homogeneous absorption
% For example, optical pumping and electromagnetically-induced absorption rely on transfer of population or coherence for enhancing the absorption
in degenerate multilevel systems
% by engaging a larger fraction of the total dipole strength
\cite{HapperRMP1972,Akulshin1999EIA,Yudin1999TOC,Goren2003EIA,TilchinPRA2011}.
These mechanisms circumvent the multilevel structure of the transition and thus can only increase the absorption up to that originating from the strongest transition in the manifold \cite{Yudin1999TOC}. In contrast, our recovery mechanism applies even when the strongest transition is directly used from the onset, as in the experiments we report here.

\emph{Recovery of absorption for atomic vapor.}---
We experimentally study the inhomogeneous compensation mechanism using $^{87}$Rb vapor at $33-42~^\circ$C.
First, we employ an $N$-type system \cite{Yudin1999TOC,Kim2010EIA,Adams2009Nsystem,Whiting2015EIA},
as depicted in Fig.~\ref{fig_spectrum}, in a 75-mm-long cell of natural abundance Rb.
The probe beam waist is 375 $\mathrm{\mu m}$, while the coupling and recovery beams are $\sim4$ times larger. We use a ring-shaped beam for Zeeman optical pumping. The probe propagates inside the dark center of the pump beam, thus avoiding undesired light shifts from the pump. Finite transit-time of atoms crossing the probe beam and noise in the phase-lock between the lasers govern the bare width of the two-photon transition $\gams=0.35$~MHz. The excited states decay rates are $\gamma=2.875$~MHz and $\gamr=3.033$~MHz.

% Figure 3
\begin{figure}[t] % [b] is a position specifier, sets to the bottom of the page.
\includegraphics[scale=0.25,trim=0.cm 0.1cm 0.cm 0.0cm,clip=true]{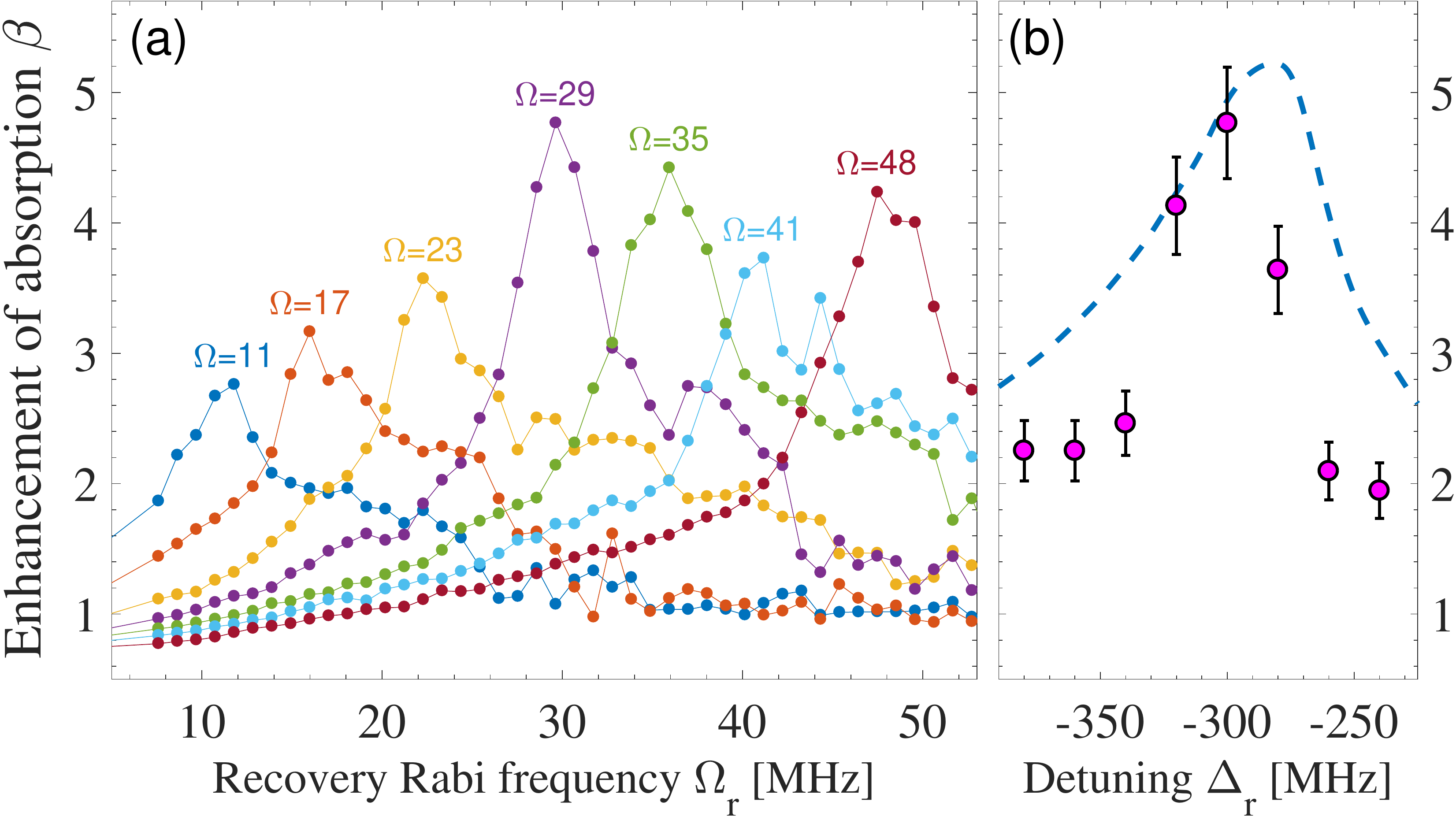}
	\caption{Enhancement of absorption $\enh$, defined with respect to the inhomogeneous limit. (a) Measured enhancement for several coupling-field intensities ($\Dc=-270$ MHz, and $\Dr$ is optimized to maximize $\enh$).
	(b) Measured enhancement (circles) versus the detuning $\Dr$ ($\Dc=-270$ MHz, and $\Omc$, $\Omr$ are optimized to maximize $\enh$). The maximal enhancement is obtained at $\Omr\approx \Omc$ [in (a)] and at $\Dr\approx\Dc$ [in (b)], which together satisfy the compensation condition \eqref{eqn:compensation_condition}. The data is consistent with full calculations [dashed blue in (b)], which include the non-idealities of the experiment (see main text).
	}
	\label{fig3a} % label is useful for referring to the figure in the text.
\end{figure}

Figure~\ref{fig_spectrum} shows absorption spectra for a bare three-level system (blue) and with the additional recovery field (green). Once the recovery field is turned on, the two-photon absorption line becomes narrow and high, considerably surpassing the inhomogeneous limit.

In order to characterize the mechanism, we measure multiple spectra at a fixed coupling-field detuning of $\Dc=-270$ MHz, and scan the intensities of the coupling and recovery fields, as well as the detuning $\Dr$ of the recovery field. We extract the enhancement $\enh$ for each spectrum from the ratio between the peak absorption at the two-photon and one-photon resonances.
% \footnote{We adopt this conservative definition (comparing the peak adsorptions when the recovery field is present) in order to account for the small pumping effect of the recovery field. For example, the compensated spectrum shown in green in Fig.~\ref{fig_spectrum} yields an enhancement factor of $\enh=4.8\pm 0.4$; if the compensated two-photon peak is compared to the one-photon peak of the uncompensated spectrum, shown in blue, the ratio would be $5.3\pm0.4$}
Figure \ref{fig3a}(a) shows the dependence of $\enh$ on $\Omr$ for several values of $\Omc$, where for each $\Omc$ we choose $\Dr$ which maximizes the enhancement. We find, as expected from condition \eqref{eqn:compensation_condition}, that the maximal enhancement is obtained when $\Omr \approx \Omc$.  In Fig.~\ref{fig3a}(b), we show the dependence of the maximal enhancement on $\Dr$. The enhancement is maximal at $\Dr \approx \Dc$ and is sensitive to deviations of order $\pm10$~MHz around this condition. This sensitivity agrees well with the requirement that Eq.~\eqref{eqn:compensation_condition} is satisfied up to the homogeneous linewidth of the two-photon resonance.

\begin{figure}[t] % [b] is a position specifier, sets to the bottom of the page.
\includegraphics[scale=0.25,trim=0.cm 0.1cm 0.cm 0.cm,clip=true]{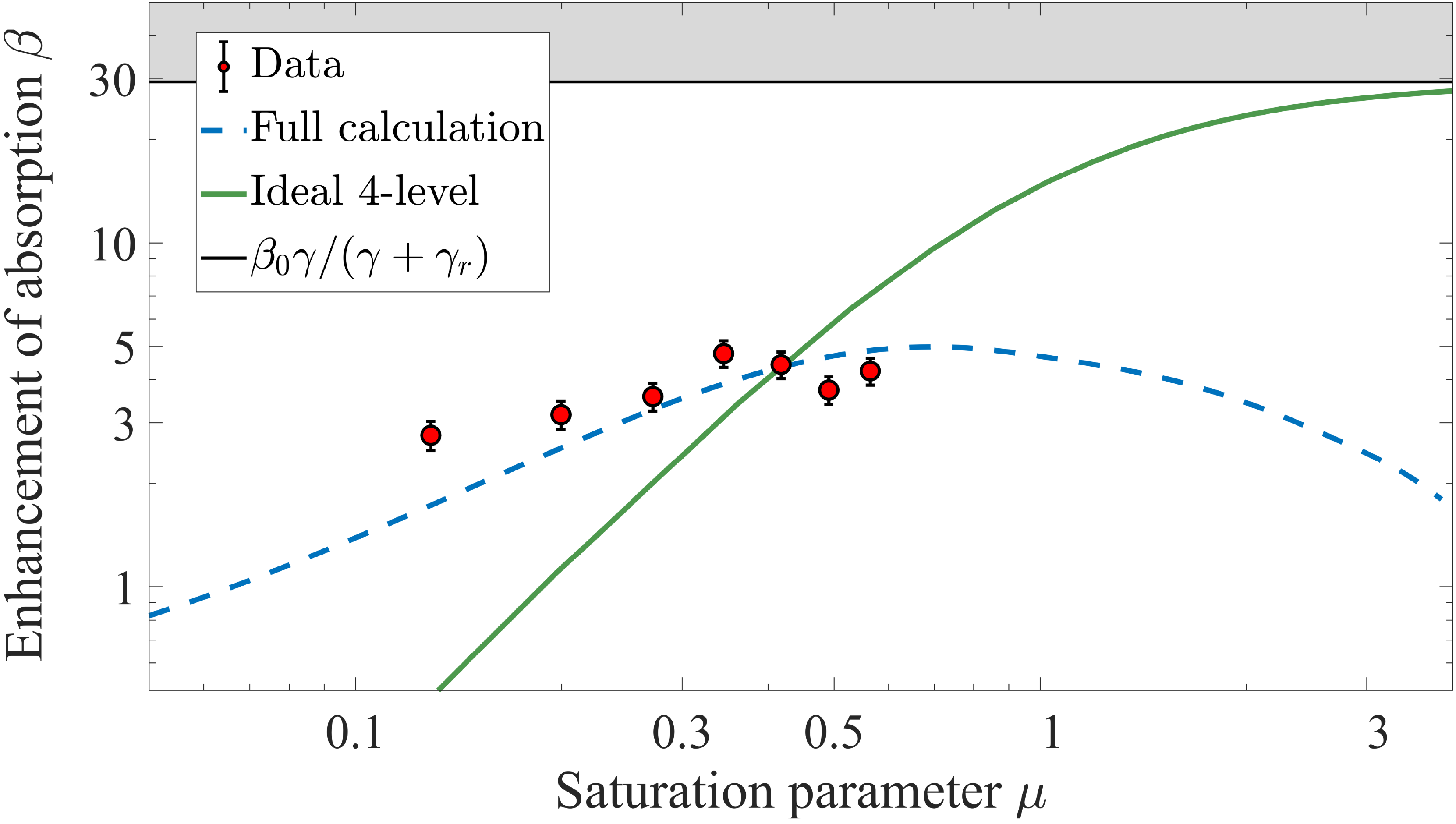}	%  trim={<left> <lower> <right> <upper>}, scale was 0.25
	\caption{Enhancement of absorption $\enh$ versus the coupling Rabi-frequency $\Omc$ in terms of  $\mu=\Omc\sqrt{(\gamma + \gamr)/[(\Dc^2+\sigma^2)\gams]}$. The experimental data (red circles) agree with the full calculation (dashed blue), which includes the non-idealities of the experiment; here $\Dc=-270$~MHz, and $\Omr$ and $\Dr$ are chosen to maximize $\enh$. The non-idealities limit the enhancement to $\enh\approx 5$. 
	In a calculation of an ideal far-detuned ($\Dc=-5$~GHz) four-level system (solid green), the enhancement approaches $\maxenh\gamma/(\gamma+\gamr)$ ($\maxenh=60$ and $\gamma\approx\gamr$), as described by Eq.~(\ref{eqn:saturation}). Reaching this upper limit (solid black) requires strong coupling and recovery fields $\mu^2\gg1$.}\label{fig4a} 
\end{figure}

Figure~\ref{fig4a} presents the dependence of $\enh$ on the intensity of the coupling and recovery fields. Ideally, according to Eq.~\eqref{eqn:saturation} with $\mu^2\gg 1$ and $\gamma\approx\gamr$, the enhancement  should saturate at $\enh\to\maxenh/2\approx 30$, as indeed obtained in a calculation of the simple four-level system (green curve).
% However, several aspects of the experimental system render it non-ideal and limit the achievable enhancement. First, the excited level has a hyperfine structure, with a splitting not much larger than the inhomogeneous (Doppler) width $\sigma$. This leads to attenuation of the Raman transition we employed (where the probe and coupling fields have orthogonal circular-polarizations) and also introduces additional velocity-dependent light-shifts.  
% Second, since $\Dc$ and $\Dr$ are not much larger than $\sigma$, there is a fraction of the ensemble that is resonant with the recovery field. Mostly, this leads to some depletion of population in $\ket{g}$ that is excited to $\ket{r}$. 
% Third, the finite beam waists of our beams, required due to limited available laser power, lead to nonuniform intensities of the coupling and recovery fields experienced by the probe and hence to nonuniform light shifts. Due to these factors, there is an optimal coupling-field power, above which the enhancement decreases.
In our experimental system, the main deviations from ideal conditions are an additional hyperfine level in the excited manifold and the nonuniform intensity of our beams due to their finite sizes.
A full calculation accounting for these with no fit parameters
% , which includes the additional hyperfine state and averages over the nonuniform intensity,
quantitatively reproduces the measured trends and the maximal enhancement [dashed blue lines in Figs.~\ref{fig3a}(b) and \ref{fig4a}].

\begin{figure}[t] % [b] is a position specifier, sets to the bottom of the page.
\includegraphics[scale=0.25,trim=0.cm 7.5cm 0.cm 0.1cm,clip=true]{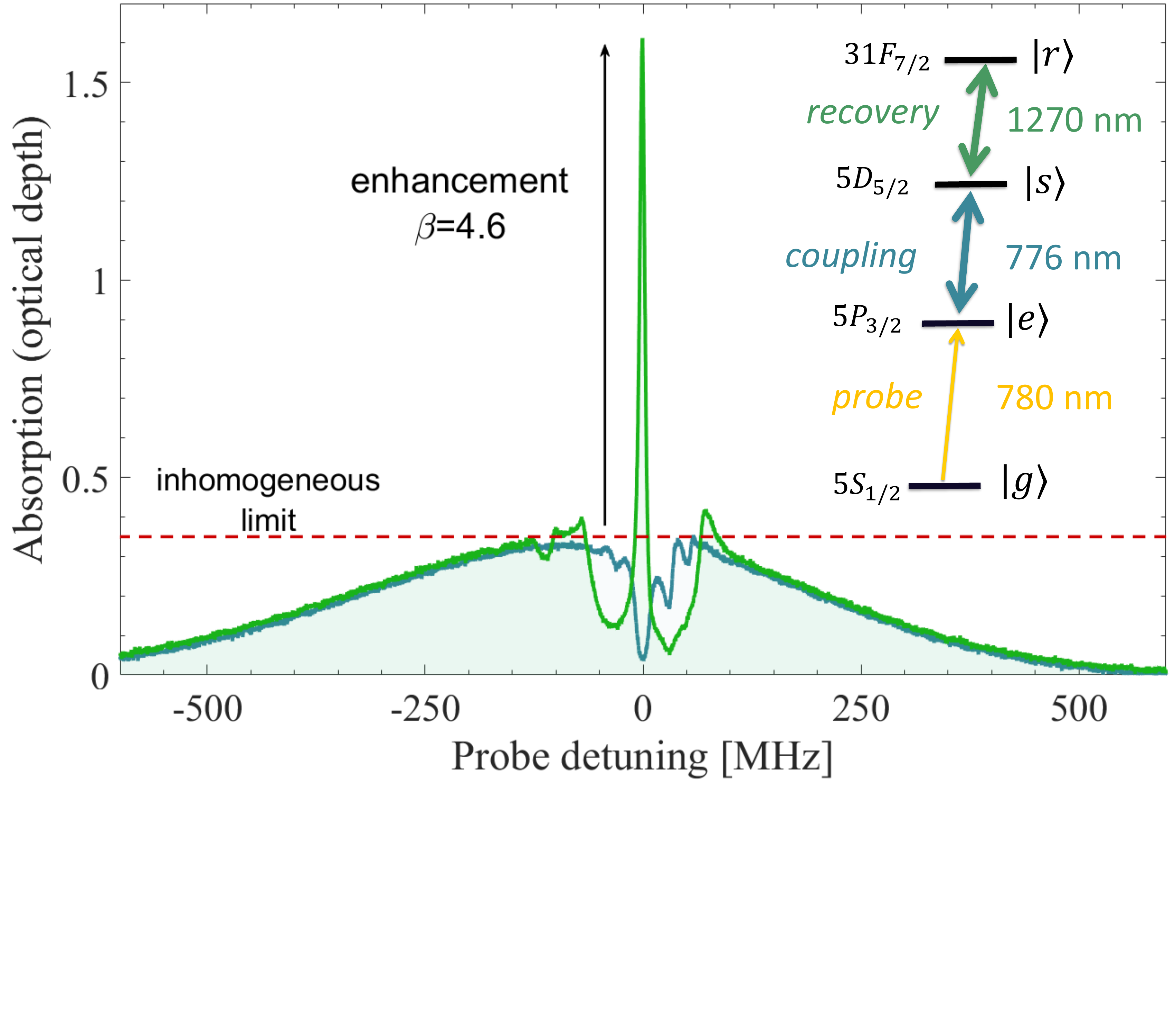} %  trim={<left> <lower> <right> <upper>
	\caption{Inhomogeneous compensation for a ladder system with a Rydberg level. The absorption spectrum in the absence of the recovery field (blue) is bounded by the inhomogeneous limit (dashed red).
	The addition of the recovery field significantly enhances the absorption of the probe (green) above this limit. In this measurement, the driving lasers with $\Omc=55$~MHz and $\Omr=45$~MHz are tuned to resonance.
	}
	\label{fig_Rydberg} % label is useful for referring to the figure in the text.
\end{figure}

To confirm the generality of the recovery mechanism, we also apply it to
% with Rydberg excitations, where a large absorption cross-section is highly desirable for quantum nonlinear effects. To this end, we employ 
a four-level ladder scheme of $^{87}$Rb in an isotopically-pure 5-mm-long cell (Fig.~\ref{fig_Rydberg}).
The probe beam, with a waist of $85~\mathrm{\mu m}$, counterpropagates the coupling and recovery beams, with waists of $750~\mathrm{\mu m}$ and $220~\mathrm{\mu m}$, respectively.
Here, the decay rates $\gams\approx 1.25$~MHz and $\gamr\approx 1$~MHz include the radiative lifetimes of the $5D$ and $31F$ levels, respectively.
In this ladder system, the two-photon transition $\ket{g}\leftrightarrow \ket{s}$ exhibits a small residual Doppler broadening $\sigma_2=1$~MHz, satisfying $\sigma_2\ll\sigma$. As shown in Fig.~\ref{fig_Rydberg},
the recovery mechanism applies equally well, yielding
a substantial enhancement of the absorption $\enh=4.6\pm0.3$ above the inhomogeneous limit.

% Figure 4

\emph{Conclusions.}---
We presented a limit to the absorption in an inhomogeneous medium and showed how light shifts arising from strong driving fields can compensate for the inhomogeneity, circumvent this limit, and recover a substantial part of the absorption cross-section. We experimentally studied the recovery mechanism in atomic vapor and demonstrated an enhancement of the absorption cross-section up to $\enh=4.8\pm0.4$ times higher than the inhomogeneous limit. The attainable recovery is limited by the available intensities of the driving fields (via the saturation parameter $\mu$) and by 
the ratio $\gamma/\gamr$ between the homogeneous widths of the employed transitions. Ideal conditions $\mu, \gamma/\gamr \gg 1$ allow for full recovery of the homogeneous absorption.

The mechanism is general and can be applied to various inhomogeneous systems in the gas and solid phases. Our experiments show that both spin and orbital transitions can be utilized (in either $N$-type or ladder-type configurations), with the recovery field coupled to either populated or unpopulated states,
and that the mechanism prevails even for small, nonzero, two-photon inhomogeneous broadening $\sigma_2\ll\sigma$.
The more extreme case, when $\sigma_2$ approaches $\sigma$, are beyond of the scope of this work \cite{finkelstein2019twocolor}.
The key requirement is that the inhomogeneous shifts $\delta$ and $\delr$ are correlated, such that condition \eqref{eqn:compensation_condition} can be satisfied.
% Finally, we demonstrate that the recovery mechanism is still applicable even
% {\blue with finite two-photon inhomogeneous broadening $\sigma_2$,} provided that $\sigma_2\ll\sigma$. The more extreme cases, when $\sigma_2$ approaches $\sigma$, are beyond the scope of this paper.

% FIGURE 5 

The enhancement of the effective absorption cross-section is essential for single-photon nonlinearities via the Rydberg blockade \cite{gorshkov2011blockade}. In this context, the ladder system employed here attracted much recent attention \cite{wade2018terahertz,Pfau2018SinglePhoton,Mohapatra2018rydberg}.
Here most crucial is the optical depth of the $\sim 10~\mathrm{\mu m }^3$ blockade volume, termed $\mathrm{OD}_{\mathrm{B}}$.
% This is manifested by, \emph{e.g.}, the intensity auto-correlation function $g^{(2)}(0) \sim  \exp{(-\ODb)}$ \cite{peyronel2012dissipative}.
In cold atoms, $\ODb>1$ is routinely achieved \cite{Lukin2018fidelity,Durr2018gate,hofferberth2018subtraction,adams2017contactless}, whereas in hot atoms, the inhomogeneous broadening severely reduces $\ODb$, making it extremely challenging to realize the strong blockade regime. Our recovery mechanism can alleviate this deficiency, leading to better scalability and higher fidelity of future quantum technology in room-temperature media.

\begin{acknowledgments}
We thank C.~Avinadav, L.~Drori, R.~Shaham, and O.~Katz for helpful discussions. We acknowledge financial support by the Israel Science Foundation and ICORE, the European Research Council starting investigator grant Q-PHOTONICS 678674, the Pazy Foundation, the Minerva Foundation with funding from the Federal German Ministry for Education and Research, and the Laboratory in Memory of Leon and Blacky Broder.
\end{acknowledgments}

% \bibliographystyle{apsrev4-1}
% \bibliographystyle{plain}
%\bibliographystyle{unsrt}
% \bibliographystyle{abbrv}
% \printbibliography

\bibliography{comp_library1.bib} % This is the bib

\begin{thebibliography}{47}%
\makeatletter
\providecommand \@ifxundefined [1]{%
 \@ifx{#1\undefined}
}%
\providecommand \@ifnum [1]{%
 \ifnum #1\expandafter \@firstoftwo
 \else \expandafter \@secondoftwo
 \fi
}%
\providecommand \@ifx [1]{%
 \ifx #1\expandafter \@firstoftwo
 \else \expandafter \@secondoftwo
 \fi
}%
\providecommand \natexlab [1]{#1}%
\providecommand \enquote  [1]{``#1''}%
\providecommand \bibnamefont  [1]{#1}%
\providecommand \bibfnamefont [1]{#1}%
\providecommand \citenamefont [1]{#1}%
\providecommand \href@noop [0]{\@secondoftwo}%
\providecommand \href [0]{\begingroup \@sanitize@url \@href}%
\providecommand \@href[1]{\@@startlink{#1}\@@href}%
\providecommand \@@href[1]{\endgroup#1\@@endlink}%
\providecommand \@sanitize@url [0]{\catcode `\\12\catcode `\$12\catcode
  `\&12\catcode `\#12\catcode `\^12\catcode `\_12\catcode `\%12\relax}%
\providecommand \@@startlink[1]{}%
\providecommand \@@endlink[0]{}%
\providecommand \url  [0]{\begingroup\@sanitize@url \@url }%
\providecommand \@url [1]{\endgroup\@href {#1}{\urlprefix }}%
\providecommand \urlprefix  [0]{URL }%
\providecommand \Eprint [0]{\href }%
\providecommand \doibase [0]{http://dx.doi.org/}%
\providecommand \selectlanguage [0]{\@gobble}%
\providecommand \bibinfo  [0]{\@secondoftwo}%
\providecommand \bibfield  [0]{\@secondoftwo}%
\providecommand \translation [1]{[#1]}%
\providecommand \BibitemOpen [0]{}%
\providecommand \bibitemStop [0]{}%
\providecommand \bibitemNoStop [0]{.\EOS\space}%
\providecommand \EOS [0]{\spacefactor3000\relax}%
\providecommand \BibitemShut  [1]{\csname bibitem#1\endcsname}%
\let\auto@bib@innerbib\@empty
%</preamble>
\bibitem [{\citenamefont {Borri}\ \emph {et~al.}(2002)\citenamefont {Borri},
  \citenamefont {Langbein}, \citenamefont {Schneider}, \citenamefont {Woggon},
  \citenamefont {Sellin}, \citenamefont {Ouyang},\ and\ \citenamefont
  {Bimberg}}]{Woggon2002QDs}%
  \BibitemOpen
  \bibfield  {author} {\bibinfo {author} {\bibfnamefont {P.}~\bibnamefont
  {Borri}}, \bibinfo {author} {\bibfnamefont {W.}~\bibnamefont {Langbein}},
  \bibinfo {author} {\bibfnamefont {S.}~\bibnamefont {Schneider}}, \bibinfo
  {author} {\bibfnamefont {U.}~\bibnamefont {Woggon}}, \bibinfo {author}
  {\bibfnamefont {R.~L.}\ \bibnamefont {Sellin}}, \bibinfo {author}
  {\bibfnamefont {D.}~\bibnamefont {Ouyang}}, \ and\ \bibinfo {author}
  {\bibfnamefont {D.}~\bibnamefont {Bimberg}},\ }\bibfield  {title} {\enquote
  {\bibinfo {title} {Rabi oscillations in the excitonic ground-state transition
  of {InGaAs} quantum dots},}\ }\href {\doibase 10.1103/PhysRevB.66.081306}
  {\bibfield  {journal} {\bibinfo  {journal} {Phys. Rev. B}\ }\textbf {\bibinfo
  {volume} {66}},\ \bibinfo {pages} {081306} (\bibinfo {year}
  {2002})}\BibitemShut {NoStop}%
\bibitem [{\citenamefont {Dutt}\ \emph {et~al.}(2005)\citenamefont {Dutt},
  \citenamefont {Cheng}, \citenamefont {Li}, \citenamefont {Xu}, \citenamefont
  {Li}, \citenamefont {Berman}, \citenamefont {Steel}, \citenamefont {Bracker},
  \citenamefont {Gammon}, \citenamefont {Economou}, \citenamefont {Liu},\ and\
  \citenamefont {Sham}}]{Steel2005QDsRaman}%
  \BibitemOpen
  \bibfield  {author} {\bibinfo {author} {\bibfnamefont {M.~V.~Gurudev}\
  \bibnamefont {Dutt}}, \bibinfo {author} {\bibfnamefont {Jun}\ \bibnamefont
  {Cheng}}, \bibinfo {author} {\bibfnamefont {Bo}~\bibnamefont {Li}}, \bibinfo
  {author} {\bibfnamefont {Xiaodong}\ \bibnamefont {Xu}}, \bibinfo {author}
  {\bibfnamefont {Xiaoqin}\ \bibnamefont {Li}}, \bibinfo {author}
  {\bibfnamefont {P.~R.}\ \bibnamefont {Berman}}, \bibinfo {author}
  {\bibfnamefont {D.~G.}\ \bibnamefont {Steel}}, \bibinfo {author}
  {\bibfnamefont {A.~S.}\ \bibnamefont {Bracker}}, \bibinfo {author}
  {\bibfnamefont {D.}~\bibnamefont {Gammon}}, \bibinfo {author} {\bibfnamefont
  {Sophia~E.}\ \bibnamefont {Economou}}, \bibinfo {author} {\bibfnamefont
  {Ren-Bao}\ \bibnamefont {Liu}}, \ and\ \bibinfo {author} {\bibfnamefont
  {L.~J.}\ \bibnamefont {Sham}},\ }\bibfield  {title} {\enquote {\bibinfo
  {title} {Stimulated and spontaneous optical generation of electron spin
  coherence in charged {GaAs} quantum dots},}\ }\href {\doibase
  10.1103/PhysRevLett.94.227403} {\bibfield  {journal} {\bibinfo  {journal}
  {Phys. Rev. Lett.}\ }\textbf {\bibinfo {volume} {94}},\ \bibinfo {pages}
  {227403} (\bibinfo {year} {2005})}\BibitemShut {NoStop}%
\bibitem [{\citenamefont {Gupta}\ \emph {et~al.}(1999)\citenamefont {Gupta},
  \citenamefont {Awschalom}, \citenamefont {Peng},\ and\ \citenamefont
  {Alivisatos}}]{Alivisatos1999QDs}%
  \BibitemOpen
  \bibfield  {author} {\bibinfo {author} {\bibfnamefont {J.~A.}\ \bibnamefont
  {Gupta}}, \bibinfo {author} {\bibfnamefont {D.~D.}\ \bibnamefont
  {Awschalom}}, \bibinfo {author} {\bibfnamefont {X.}~\bibnamefont {Peng}}, \
  and\ \bibinfo {author} {\bibfnamefont {A.~P.}\ \bibnamefont {Alivisatos}},\
  }\bibfield  {title} {\enquote {\bibinfo {title} {Spin coherence in
  semiconductor quantum dots},}\ }\href {\doibase 10.1103/PhysRevB.59.R10421}
  {\bibfield  {journal} {\bibinfo  {journal} {Phys. Rev. B}\ }\textbf {\bibinfo
  {volume} {59}},\ \bibinfo {pages} {R10421--R10424} (\bibinfo {year}
  {1999})}\BibitemShut {NoStop}%
\bibitem [{\citenamefont {Pingault}\ \emph {et~al.}(2014)\citenamefont
  {Pingault}, \citenamefont {Becker}, \citenamefont {Schulte}, \citenamefont
  {Arend}, \citenamefont {Hepp}, \citenamefont {Godde}, \citenamefont
  {Tartakovskii}, \citenamefont {Markham}, \citenamefont {Becher},\ and\
  \citenamefont {Atat\"ure}}]{atature2014single}%
  \BibitemOpen
  \bibfield  {author} {\bibinfo {author} {\bibfnamefont {Benjamin}\
  \bibnamefont {Pingault}}, \bibinfo {author} {\bibfnamefont {Jonas~N.}\
  \bibnamefont {Becker}}, \bibinfo {author} {\bibfnamefont {Carsten H.~H.}\
  \bibnamefont {Schulte}}, \bibinfo {author} {\bibfnamefont {Carsten}\
  \bibnamefont {Arend}}, \bibinfo {author} {\bibfnamefont {Christian}\
  \bibnamefont {Hepp}}, \bibinfo {author} {\bibfnamefont {Tillmann}\
  \bibnamefont {Godde}}, \bibinfo {author} {\bibfnamefont {Alexander~I.}\
  \bibnamefont {Tartakovskii}}, \bibinfo {author} {\bibfnamefont {Matthew}\
  \bibnamefont {Markham}}, \bibinfo {author} {\bibfnamefont {Christoph}\
  \bibnamefont {Becher}}, \ and\ \bibinfo {author} {\bibfnamefont {Mete}\
  \bibnamefont {Atat\"ure}},\ }\bibfield  {title} {\enquote {\bibinfo {title}
  {All-optical formation of coherent dark states of silicon-vacancy spins in
  diamond},}\ }\href {\doibase 10.1103/PhysRevLett.113.263601} {\bibfield
  {journal} {\bibinfo  {journal} {Phys. Rev. Lett.}\ }\textbf {\bibinfo
  {volume} {113}},\ \bibinfo {pages} {263601} (\bibinfo {year}
  {2014})}\BibitemShut {NoStop}%
\bibitem [{\citenamefont {Rogers}\ \emph {et~al.}(2014)\citenamefont {Rogers},
  \citenamefont {Jahnke}, \citenamefont {Metsch}, \citenamefont {Sipahigil},
  \citenamefont {Binder}, \citenamefont {Teraji}, \citenamefont {Sumiya},
  \citenamefont {Isoya}, \citenamefont {Lukin}, \citenamefont {Hemmer},\ and\
  \citenamefont {Jelezko}}]{jelezko2014single}%
  \BibitemOpen
  \bibfield  {author} {\bibinfo {author} {\bibfnamefont {Lachlan~J.}\
  \bibnamefont {Rogers}}, \bibinfo {author} {\bibfnamefont {Kay~D.}\
  \bibnamefont {Jahnke}}, \bibinfo {author} {\bibfnamefont {Mathias~H.}\
  \bibnamefont {Metsch}}, \bibinfo {author} {\bibfnamefont {Alp}\ \bibnamefont
  {Sipahigil}}, \bibinfo {author} {\bibfnamefont {Jan~M.}\ \bibnamefont
  {Binder}}, \bibinfo {author} {\bibfnamefont {Tokuyuki}\ \bibnamefont
  {Teraji}}, \bibinfo {author} {\bibfnamefont {Hitoshi}\ \bibnamefont
  {Sumiya}}, \bibinfo {author} {\bibfnamefont {Junichi}\ \bibnamefont {Isoya}},
  \bibinfo {author} {\bibfnamefont {Mikhail~D.}\ \bibnamefont {Lukin}},
  \bibinfo {author} {\bibfnamefont {Philip}\ \bibnamefont {Hemmer}}, \ and\
  \bibinfo {author} {\bibfnamefont {Fedor}\ \bibnamefont {Jelezko}},\
  }\bibfield  {title} {\enquote {\bibinfo {title} {All-optical initialization,
  readout, and coherent preparation of single silicon-vacancy spins in
  diamond},}\ }\href {\doibase 10.1103/PhysRevLett.113.263602} {\bibfield
  {journal} {\bibinfo  {journal} {Phys. Rev. Lett.}\ }\textbf {\bibinfo
  {volume} {113}},\ \bibinfo {pages} {263602} (\bibinfo {year}
  {2014})}\BibitemShut {NoStop}%
\bibitem [{\citenamefont {Lauritzen}\ \emph {et~al.}(2010)\citenamefont
  {Lauritzen}, \citenamefont {Min\'a\ifmmode~\check{r}\else \v{r}\fi{}},
  \citenamefont {de~Riedmatten}, \citenamefont {Afzelius}, \citenamefont
  {Sangouard}, \citenamefont {Simon},\ and\ \citenamefont
  {Gisin}}]{Gisin2010telecom}%
  \BibitemOpen
  \bibfield  {author} {\bibinfo {author} {\bibfnamefont {Bj\"orn}\ \bibnamefont
  {Lauritzen}}, \bibinfo {author} {\bibfnamefont {Ji\ifmmode
  \check{r}\else~\v{r}\fi{}\'{\i}}\ \bibnamefont {Min\'a\ifmmode~\check{r}\else
  \v{r}\fi{}}}, \bibinfo {author} {\bibfnamefont {Hugues}\ \bibnamefont
  {de~Riedmatten}}, \bibinfo {author} {\bibfnamefont {Mikael}\ \bibnamefont
  {Afzelius}}, \bibinfo {author} {\bibfnamefont {Nicolas}\ \bibnamefont
  {Sangouard}}, \bibinfo {author} {\bibfnamefont {Christoph}\ \bibnamefont
  {Simon}}, \ and\ \bibinfo {author} {\bibfnamefont {Nicolas}\ \bibnamefont
  {Gisin}},\ }\bibfield  {title} {\enquote {\bibinfo {title}
  {Telecommunication-wavelength solid-state memory at the single photon
  level},}\ }\href {\doibase 10.1103/PhysRevLett.104.080502} {\bibfield
  {journal} {\bibinfo  {journal} {Phys. Rev. Lett.}\ }\textbf {\bibinfo
  {volume} {104}},\ \bibinfo {pages} {080502} (\bibinfo {year}
  {2010})}\BibitemShut {NoStop}%
\bibitem [{\citenamefont {Simon}\ \emph {et~al.}(2010)\citenamefont {Simon},
  \citenamefont {Afzelius}, \citenamefont {Appel}, \citenamefont {Boyer de~la
  Giroday}, \citenamefont {Dewhurst}, \citenamefont {Gisin}, \citenamefont
  {Hu}, \citenamefont {Jelezko}, \citenamefont {Kr{\"o}ll}, \citenamefont
  {M{\"u}ller}, \citenamefont {Nunn}, \citenamefont {Polzik}, \citenamefont
  {Rarity}, \citenamefont {De~Riedmatten}, \citenamefont {Rosenfeld},
  \citenamefont {Shields}, \citenamefont {Sk{\"o}ld}, \citenamefont
  {Stevenson}, \citenamefont {Thew}, \citenamefont {Walmsley}, \citenamefont
  {Weber}, \citenamefont {Weinfurter}, \citenamefont {Wrachtrup},\ and\
  \citenamefont {Young}}]{Qreview2010}%
  \BibitemOpen
  \bibfield  {author} {\bibinfo {author} {\bibfnamefont {C.}~\bibnamefont
  {Simon}}, \bibinfo {author} {\bibfnamefont {M.}~\bibnamefont {Afzelius}},
  \bibinfo {author} {\bibfnamefont {J.}~\bibnamefont {Appel}}, \bibinfo
  {author} {\bibfnamefont {A.}~\bibnamefont {Boyer de~la Giroday}}, \bibinfo
  {author} {\bibfnamefont {S.~J.}\ \bibnamefont {Dewhurst}}, \bibinfo {author}
  {\bibfnamefont {N.}~\bibnamefont {Gisin}}, \bibinfo {author} {\bibfnamefont
  {C.~Y.}\ \bibnamefont {Hu}}, \bibinfo {author} {\bibfnamefont
  {F.}~\bibnamefont {Jelezko}}, \bibinfo {author} {\bibfnamefont
  {S.}~\bibnamefont {Kr{\"o}ll}}, \bibinfo {author} {\bibfnamefont {J.~H.}\
  \bibnamefont {M{\"u}ller}}, \bibinfo {author} {\bibfnamefont
  {J.}~\bibnamefont {Nunn}}, \bibinfo {author} {\bibfnamefont {E.~S.}\
  \bibnamefont {Polzik}}, \bibinfo {author} {\bibfnamefont {J.~G.}\
  \bibnamefont {Rarity}}, \bibinfo {author} {\bibfnamefont {H.}~\bibnamefont
  {De~Riedmatten}}, \bibinfo {author} {\bibfnamefont {W.}~\bibnamefont
  {Rosenfeld}}, \bibinfo {author} {\bibfnamefont {A.~J.}\ \bibnamefont
  {Shields}}, \bibinfo {author} {\bibfnamefont {N.}~\bibnamefont {Sk{\"o}ld}},
  \bibinfo {author} {\bibfnamefont {R.~M.}\ \bibnamefont {Stevenson}}, \bibinfo
  {author} {\bibfnamefont {R.}~\bibnamefont {Thew}}, \bibinfo {author}
  {\bibfnamefont {I.~A.}\ \bibnamefont {Walmsley}}, \bibinfo {author}
  {\bibfnamefont {M.~C.}\ \bibnamefont {Weber}}, \bibinfo {author}
  {\bibfnamefont {H.}~\bibnamefont {Weinfurter}}, \bibinfo {author}
  {\bibfnamefont {J.}~\bibnamefont {Wrachtrup}}, \ and\ \bibinfo {author}
  {\bibfnamefont {R.~J.}\ \bibnamefont {Young}},\ }\bibfield  {title} {\enquote
  {\bibinfo {title} {Quantum memories},}\ }\href {\doibase
  10.1140/epjd/e2010-00103-y} {\bibfield  {journal} {\bibinfo  {journal} {The
  European Physical Journal D}\ }\textbf {\bibinfo {volume} {58}},\ \bibinfo
  {pages} {1--22} (\bibinfo {year} {2010})}\BibitemShut {NoStop}%
\bibitem [{\citenamefont {Knappe}\ \emph {et~al.}(2005)\citenamefont {Knappe},
  \citenamefont {Schwindt}, \citenamefont {Shah}, \citenamefont {Hollberg},
  \citenamefont {Kitching}, \citenamefont {Liew},\ and\ \citenamefont
  {Moreland}}]{Kitching2005}%
  \BibitemOpen
  \bibfield  {author} {\bibinfo {author} {\bibfnamefont {S.}~\bibnamefont
  {Knappe}}, \bibinfo {author} {\bibfnamefont {P.D.D.}\ \bibnamefont
  {Schwindt}}, \bibinfo {author} {\bibfnamefont {V.}~\bibnamefont {Shah}},
  \bibinfo {author} {\bibfnamefont {L.}~\bibnamefont {Hollberg}}, \bibinfo
  {author} {\bibfnamefont {J.}~\bibnamefont {Kitching}}, \bibinfo {author}
  {\bibfnamefont {L.}~\bibnamefont {Liew}}, \ and\ \bibinfo {author}
  {\bibfnamefont {J.}~\bibnamefont {Moreland}},\ }\bibfield  {title} {\enquote
  {\bibinfo {title} {A chip-scale atomic clock based on $^{87}${Rb} with
  improved frequency stability},}\ }\href {\doibase 10.1364/OPEX.13.001249}
  {\bibfield  {journal} {\bibinfo  {journal} {Opt. Express}\ }\textbf {\bibinfo
  {volume} {13}},\ \bibinfo {pages} {1249--1253} (\bibinfo {year}
  {2005})}\BibitemShut {NoStop}%
\bibitem [{\citenamefont {Carr}\ \emph {et~al.}(2012)\citenamefont {Carr},
  \citenamefont {Tanasittikosol}, \citenamefont {Sargsyan}, \citenamefont
  {Sarkisyan}, \citenamefont {Adams},\ and\ \citenamefont
  {Weatherill}}]{carr2012three}%
  \BibitemOpen
  \bibfield  {author} {\bibinfo {author} {\bibfnamefont {Christopher}\
  \bibnamefont {Carr}}, \bibinfo {author} {\bibfnamefont {Monsit}\ \bibnamefont
  {Tanasittikosol}}, \bibinfo {author} {\bibfnamefont {Armen}\ \bibnamefont
  {Sargsyan}}, \bibinfo {author} {\bibfnamefont {David}\ \bibnamefont
  {Sarkisyan}}, \bibinfo {author} {\bibfnamefont {Charles~S}\ \bibnamefont
  {Adams}}, \ and\ \bibinfo {author} {\bibfnamefont {Kevin~J}\ \bibnamefont
  {Weatherill}},\ }\bibfield  {title} {\enquote {\bibinfo {title} {Three-photon
  electromagnetically induced transparency using {Rydberg} states},}\
  }\href@noop {} {\bibfield  {journal} {\bibinfo  {journal} {Optics letters}\
  }\textbf {\bibinfo {volume} {37}},\ \bibinfo {pages} {3858--3860} (\bibinfo
  {year} {2012})}\BibitemShut {NoStop}%
\bibitem [{\citenamefont {Ripka}\ \emph {et~al.}(2018)\citenamefont {Ripka},
  \citenamefont {K{\"u}bler}, \citenamefont {L{\"o}w},\ and\ \citenamefont
  {Pfau}}]{Pfau2018SinglePhoton}%
  \BibitemOpen
  \bibfield  {author} {\bibinfo {author} {\bibfnamefont {Fabian}\ \bibnamefont
  {Ripka}}, \bibinfo {author} {\bibfnamefont {Harald}\ \bibnamefont
  {K{\"u}bler}}, \bibinfo {author} {\bibfnamefont {Robert}\ \bibnamefont
  {L{\"o}w}}, \ and\ \bibinfo {author} {\bibfnamefont {Tilman}\ \bibnamefont
  {Pfau}},\ }\bibfield  {title} {\enquote {\bibinfo {title} {A room-temperature
  single-photon source based on strongly interacting {Rydberg} atoms},}\ }\href
  {\doibase 10.1126/science.aau1949} {\bibfield  {journal} {\bibinfo  {journal}
  {Science}\ }\textbf {\bibinfo {volume} {362}},\ \bibinfo {pages} {446--449}
  (\bibinfo {year} {2018})}\BibitemShut {NoStop}%
\bibitem [{\citenamefont {Demtr{\"o}der}(2015)}]{demtroder2015laser}%
  \BibitemOpen
  \bibfield  {author} {\bibinfo {author} {\bibfnamefont {Wolfgang}\
  \bibnamefont {Demtr{\"o}der}},\ }\href@noop {} {\emph {\bibinfo {title}
  {Laser Spectroscopy 2: Experimental Techniques}}}\ (\bibinfo  {publisher}
  {Springer},\ \bibinfo {year} {2015})\BibitemShut {NoStop}%
\bibitem [{\citenamefont {Kurnit}\ \emph {et~al.}(1964)\citenamefont {Kurnit},
  \citenamefont {Abella},\ and\ \citenamefont {Hartmann}}]{Kurnit1964echo}%
  \BibitemOpen
  \bibfield  {author} {\bibinfo {author} {\bibfnamefont {N.~A.}\ \bibnamefont
  {Kurnit}}, \bibinfo {author} {\bibfnamefont {I.~D.}\ \bibnamefont {Abella}},
  \ and\ \bibinfo {author} {\bibfnamefont {S.~R.}\ \bibnamefont {Hartmann}},\
  }\bibfield  {title} {\enquote {\bibinfo {title} {Observation of a photon
  echo},}\ }\href {\doibase 10.1103/PhysRevLett.13.567} {\bibfield  {journal}
  {\bibinfo  {journal} {Phys. Rev. Lett.}\ }\textbf {\bibinfo {volume} {13}},\
  \bibinfo {pages} {567--568} (\bibinfo {year} {1964})}\BibitemShut {NoStop}%
\bibitem [{\citenamefont {Kraus}\ \emph {et~al.}(2006)\citenamefont {Kraus},
  \citenamefont {Tittel}, \citenamefont {Gisin}, \citenamefont {Nilsson},
  \citenamefont {Kr\"oll},\ and\ \citenamefont {Cirac}}]{Kraus2006CRIB}%
  \BibitemOpen
  \bibfield  {author} {\bibinfo {author} {\bibfnamefont {B.}~\bibnamefont
  {Kraus}}, \bibinfo {author} {\bibfnamefont {W.}~\bibnamefont {Tittel}},
  \bibinfo {author} {\bibfnamefont {N.}~\bibnamefont {Gisin}}, \bibinfo
  {author} {\bibfnamefont {M.}~\bibnamefont {Nilsson}}, \bibinfo {author}
  {\bibfnamefont {S.}~\bibnamefont {Kr\"oll}}, \ and\ \bibinfo {author}
  {\bibfnamefont {J.~I.}\ \bibnamefont {Cirac}},\ }\bibfield  {title} {\enquote
  {\bibinfo {title} {Quantum memory for nonstationary light fields based on
  controlled reversible inhomogeneous broadening},}\ }\href {\doibase
  10.1103/PhysRevA.73.020302} {\bibfield  {journal} {\bibinfo  {journal} {Phys.
  Rev. A}\ }\textbf {\bibinfo {volume} {73}},\ \bibinfo {pages} {020302}
  (\bibinfo {year} {2006})}\BibitemShut {NoStop}%
\bibitem [{\citenamefont {Rickes}\ \emph {et~al.}(2000)\citenamefont {Rickes},
  \citenamefont {Yatsenko}, \citenamefont {Steuerwald}, \citenamefont
  {Halfmann}, \citenamefont {Shore}, \citenamefont {Vitanov},\ and\
  \citenamefont {Bergmann}}]{Bergmann2000SCRAP}%
  \BibitemOpen
  \bibfield  {author} {\bibinfo {author} {\bibfnamefont {T}~\bibnamefont
  {Rickes}}, \bibinfo {author} {\bibfnamefont {LP}~\bibnamefont {Yatsenko}},
  \bibinfo {author} {\bibfnamefont {S}~\bibnamefont {Steuerwald}}, \bibinfo
  {author} {\bibfnamefont {T}~\bibnamefont {Halfmann}}, \bibinfo {author}
  {\bibfnamefont {BW}~\bibnamefont {Shore}}, \bibinfo {author} {\bibfnamefont
  {NV}~\bibnamefont {Vitanov}}, \ and\ \bibinfo {author} {\bibfnamefont
  {K}~\bibnamefont {Bergmann}},\ }\bibfield  {title} {\enquote {\bibinfo
  {title} {Efficient adiabatic population transfer by two-photon excitation
  assisted by a laser-induced stark shift},}\ }\href@noop {} {\bibfield
  {journal} {\bibinfo  {journal} {The Journal of Chemical Physics}\ }\textbf
  {\bibinfo {volume} {113}},\ \bibinfo {pages} {534--546} (\bibinfo {year}
  {2000})}\BibitemShut {NoStop}%
\bibitem [{\citenamefont {Sch{\"o}nfeldt}\ \emph {et~al.}(2009)\citenamefont
  {Sch{\"o}nfeldt}, \citenamefont {Twamley},\ and\ \citenamefont
  {Rebi{\'c}}}]{Rebic2009SCRAP}%
  \BibitemOpen
  \bibfield  {author} {\bibinfo {author} {\bibfnamefont {Johann-Heinrich}\
  \bibnamefont {Sch{\"o}nfeldt}}, \bibinfo {author} {\bibfnamefont {Jason}\
  \bibnamefont {Twamley}}, \ and\ \bibinfo {author} {\bibfnamefont {Stojan}\
  \bibnamefont {Rebi{\'c}}},\ }\bibfield  {title} {\enquote {\bibinfo {title}
  {Optimized control of stark-shift-chirped rapid adiabatic passage in a
  $\lambda$-type three-level system},}\ }\href@noop {} {\bibfield  {journal}
  {\bibinfo  {journal} {Physical Review A}\ }\textbf {\bibinfo {volume} {80}},\
  \bibinfo {pages} {043401} (\bibinfo {year} {2009})}\BibitemShut {NoStop}%
\bibitem [{\citenamefont {Gorshkov}\ \emph {et~al.}(2007)\citenamefont
  {Gorshkov}, \citenamefont {Andr\'e}, \citenamefont {Lukin},\ and\
  \citenamefont {S\o{}rensen}}]{GorshkovIII}%
  \BibitemOpen
  \bibfield  {author} {\bibinfo {author} {\bibfnamefont {Alexey~V.}\
  \bibnamefont {Gorshkov}}, \bibinfo {author} {\bibfnamefont {Axel}\
  \bibnamefont {Andr\'e}}, \bibinfo {author} {\bibfnamefont {Mikhail~D.}\
  \bibnamefont {Lukin}}, \ and\ \bibinfo {author} {\bibfnamefont {Anders~S.}\
  \bibnamefont {S\o{}rensen}},\ }\bibfield  {title} {\enquote {\bibinfo {title}
  {Photon storage in $\ensuremath{\Lambda}$-type optically dense atomic media.
  iii. effects of inhomogeneous broadening},}\ }\href {\doibase
  10.1103/PhysRevA.76.033806} {\bibfield  {journal} {\bibinfo  {journal} {Phys.
  Rev. A}\ }\textbf {\bibinfo {volume} {76}},\ \bibinfo {pages} {033806}
  (\bibinfo {year} {2007})}\BibitemShut {NoStop}%
\bibitem [{\citenamefont {Nunn}\ \emph {et~al.}(2007)\citenamefont {Nunn},
  \citenamefont {Walmsley}, \citenamefont {Raymer}, \citenamefont {Surmacz},
  \citenamefont {Waldermann}, \citenamefont {Wang},\ and\ \citenamefont
  {Jaksch}}]{Josh2007}%
  \BibitemOpen
  \bibfield  {author} {\bibinfo {author} {\bibfnamefont {J.}~\bibnamefont
  {Nunn}}, \bibinfo {author} {\bibfnamefont {I.~A.}\ \bibnamefont {Walmsley}},
  \bibinfo {author} {\bibfnamefont {M.~G.}\ \bibnamefont {Raymer}}, \bibinfo
  {author} {\bibfnamefont {K.}~\bibnamefont {Surmacz}}, \bibinfo {author}
  {\bibfnamefont {F.~C.}\ \bibnamefont {Waldermann}}, \bibinfo {author}
  {\bibfnamefont {Z.}~\bibnamefont {Wang}}, \ and\ \bibinfo {author}
  {\bibfnamefont {D.}~\bibnamefont {Jaksch}},\ }\bibfield  {title} {\enquote
  {\bibinfo {title} {Mapping broadband single-photon wave packets into an
  atomic memory},}\ }\href {\doibase 10.1103/PhysRevA.75.011401} {\bibfield
  {journal} {\bibinfo  {journal} {Phys. Rev. A}\ }\textbf {\bibinfo {volume}
  {75}},\ \bibinfo {pages} {011401} (\bibinfo {year} {2007})}\BibitemShut
  {NoStop}%
\bibitem [{\citenamefont {Finkelstein}\ \emph {et~al.}(2018)\citenamefont
  {Finkelstein}, \citenamefont {Poem}, \citenamefont {Michel}, \citenamefont
  {Lahad},\ and\ \citenamefont {Firstenberg}}]{FLAME}%
  \BibitemOpen
  \bibfield  {author} {\bibinfo {author} {\bibfnamefont {Ran}\ \bibnamefont
  {Finkelstein}}, \bibinfo {author} {\bibfnamefont {Eilon}\ \bibnamefont
  {Poem}}, \bibinfo {author} {\bibfnamefont {Ohad}\ \bibnamefont {Michel}},
  \bibinfo {author} {\bibfnamefont {Ohr}\ \bibnamefont {Lahad}}, \ and\
  \bibinfo {author} {\bibfnamefont {Ofer}\ \bibnamefont {Firstenberg}},\
  }\bibfield  {title} {\enquote {\bibinfo {title} {Fast, noise-free memory for
  photon synchronization at room temperature},}\ }\href@noop {} {\bibfield
  {journal} {\bibinfo  {journal} {Science Advances}\ }\textbf {\bibinfo
  {volume} {4}} (\bibinfo {year} {2018})}\BibitemShut {NoStop}%
\bibitem [{\citenamefont {Gorshkov}\ \emph {et~al.}(2011)\citenamefont
  {Gorshkov}, \citenamefont {Otterbach}, \citenamefont {Fleischhauer},
  \citenamefont {Pohl},\ and\ \citenamefont {Lukin}}]{gorshkov2011blockade}%
  \BibitemOpen
  \bibfield  {author} {\bibinfo {author} {\bibfnamefont {Alexey~V}\
  \bibnamefont {Gorshkov}}, \bibinfo {author} {\bibfnamefont {Johannes}\
  \bibnamefont {Otterbach}}, \bibinfo {author} {\bibfnamefont {Michael}\
  \bibnamefont {Fleischhauer}}, \bibinfo {author} {\bibfnamefont {Thomas}\
  \bibnamefont {Pohl}}, \ and\ \bibinfo {author} {\bibfnamefont {Mikhail~D}\
  \bibnamefont {Lukin}},\ }\bibfield  {title} {\enquote {\bibinfo {title}
  {Photon-photon interactions via {Rydberg} blockade},}\ }\href@noop {}
  {\bibfield  {journal} {\bibinfo  {journal} {Physical review letters}\
  }\textbf {\bibinfo {volume} {107}},\ \bibinfo {pages} {133602} (\bibinfo
  {year} {2011})}\BibitemShut {NoStop}%
\bibitem [{\citenamefont {Firstenberg}\ \emph {et~al.}(2016)\citenamefont
  {Firstenberg}, \citenamefont {Adams},\ and\ \citenamefont
  {Hofferberth}}]{firstenberg2016review}%
  \BibitemOpen
  \bibfield  {author} {\bibinfo {author} {\bibfnamefont {Ofer}\ \bibnamefont
  {Firstenberg}}, \bibinfo {author} {\bibfnamefont {Charles~S}\ \bibnamefont
  {Adams}}, \ and\ \bibinfo {author} {\bibfnamefont {Sebastian}\ \bibnamefont
  {Hofferberth}},\ }\bibfield  {title} {\enquote {\bibinfo {title} {Nonlinear
  quantum optics mediated by {Rydberg} interactions},}\ }\href@noop {}
  {\bibfield  {journal} {\bibinfo  {journal} {Journal of Physics B: Atomic,
  Molecular and Optical Physics}\ }\textbf {\bibinfo {volume} {49}},\ \bibinfo
  {pages} {152003} (\bibinfo {year} {2016})}\BibitemShut {NoStop}%
\bibitem [{\citenamefont {L{\"o}w}\ \emph {et~al.}(2012)\citenamefont
  {L{\"o}w}, \citenamefont {Weimer}, \citenamefont {Nipper}, \citenamefont
  {Balewski}, \citenamefont {Butscher}, \citenamefont {B{\"u}chler},\ and\
  \citenamefont {Pfau}}]{Pfau2012Guide}%
  \BibitemOpen
  \bibfield  {author} {\bibinfo {author} {\bibfnamefont {Robert}\ \bibnamefont
  {L{\"o}w}}, \bibinfo {author} {\bibfnamefont {Hendrik}\ \bibnamefont
  {Weimer}}, \bibinfo {author} {\bibfnamefont {Johannes}\ \bibnamefont
  {Nipper}}, \bibinfo {author} {\bibfnamefont {Jonathan~B}\ \bibnamefont
  {Balewski}}, \bibinfo {author} {\bibfnamefont {Bj{\"o}rn}\ \bibnamefont
  {Butscher}}, \bibinfo {author} {\bibfnamefont {Hans~Peter}\ \bibnamefont
  {B{\"u}chler}}, \ and\ \bibinfo {author} {\bibfnamefont {Tilman}\
  \bibnamefont {Pfau}},\ }\bibfield  {title} {\enquote {\bibinfo {title} {An
  experimental and theoretical guide to strongly interacting {Rydberg}
  gases},}\ }\href@noop {} {\bibfield  {journal} {\bibinfo  {journal} {Journal
  of Physics B: Atomic, Molecular and Optical Physics}\ }\textbf {\bibinfo
  {volume} {45}},\ \bibinfo {pages} {113001} (\bibinfo {year}
  {2012})}\BibitemShut {NoStop}%
\bibitem [{\citenamefont {Tiarks}\ \emph {et~al.}(2018)\citenamefont {Tiarks},
  \citenamefont {Schmidt-Eberle}, \citenamefont {Stolz}, \citenamefont
  {Rempe},\ and\ \citenamefont {D{\"u}rr}}]{Durr2018gate}%
  \BibitemOpen
  \bibfield  {author} {\bibinfo {author} {\bibfnamefont {Daniel}\ \bibnamefont
  {Tiarks}}, \bibinfo {author} {\bibfnamefont {Steffen}\ \bibnamefont
  {Schmidt-Eberle}}, \bibinfo {author} {\bibfnamefont {Thomas}\ \bibnamefont
  {Stolz}}, \bibinfo {author} {\bibfnamefont {Gerhard}\ \bibnamefont {Rempe}},
  \ and\ \bibinfo {author} {\bibfnamefont {Stephan}\ \bibnamefont {D{\"u}rr}},\
  }\bibfield  {title} {\enquote {\bibinfo {title} {A photon--photon quantum
  gate based on {Rydberg} interactions},}\ }\href@noop {} {\bibfield  {journal}
  {\bibinfo  {journal} {Nature Physics}\ ,\ \bibinfo {pages} {1}} (\bibinfo
  {year} {2018})}\BibitemShut {NoStop}%
\bibitem [{\citenamefont {Baur}\ \emph {et~al.}(2014)\citenamefont {Baur},
  \citenamefont {Tiarks}, \citenamefont {Rempe},\ and\ \citenamefont
  {D\"urr}}]{Durr2014switch}%
  \BibitemOpen
  \bibfield  {author} {\bibinfo {author} {\bibfnamefont {Simon}\ \bibnamefont
  {Baur}}, \bibinfo {author} {\bibfnamefont {Daniel}\ \bibnamefont {Tiarks}},
  \bibinfo {author} {\bibfnamefont {Gerhard}\ \bibnamefont {Rempe}}, \ and\
  \bibinfo {author} {\bibfnamefont {Stephan}\ \bibnamefont {D\"urr}},\
  }\bibfield  {title} {\enquote {\bibinfo {title} {Single-photon switch based
  on rydberg blockade},}\ }\href {\doibase 10.1103/PhysRevLett.112.073901}
  {\bibfield  {journal} {\bibinfo  {journal} {Phys. Rev. Lett.}\ }\textbf
  {\bibinfo {volume} {112}},\ \bibinfo {pages} {073901} (\bibinfo {year}
  {2014})}\BibitemShut {NoStop}%
\bibitem [{\citenamefont {Gaj}\ \emph {et~al.}(2014)\citenamefont {Gaj},
  \citenamefont {Krupp}, \citenamefont {Balewski}, \citenamefont {L{\"o}w},
  \citenamefont {Hofferberth},\ and\ \citenamefont {Pfau}}]{pfau2014molecular}%
  \BibitemOpen
  \bibfield  {author} {\bibinfo {author} {\bibfnamefont {Anita}\ \bibnamefont
  {Gaj}}, \bibinfo {author} {\bibfnamefont {Alexander~T}\ \bibnamefont
  {Krupp}}, \bibinfo {author} {\bibfnamefont {Jonathan~B}\ \bibnamefont
  {Balewski}}, \bibinfo {author} {\bibfnamefont {Robert}\ \bibnamefont
  {L{\"o}w}}, \bibinfo {author} {\bibfnamefont {Sebastian}\ \bibnamefont
  {Hofferberth}}, \ and\ \bibinfo {author} {\bibfnamefont {Tilman}\
  \bibnamefont {Pfau}},\ }\bibfield  {title} {\enquote {\bibinfo {title} {From
  molecular spectra to a density shift in dense rydberg gases},}\ }\href@noop
  {} {\bibfield  {journal} {\bibinfo  {journal} {Nature communications}\
  }\textbf {\bibinfo {volume} {5}},\ \bibinfo {pages} {4546} (\bibinfo {year}
  {2014})}\BibitemShut {NoStop}%
\bibitem [{\citenamefont {Derevianko}\ \emph {et~al.}(2015)\citenamefont
  {Derevianko}, \citenamefont {K{\'o}m{\'a}r}, \citenamefont {Topcu},
  \citenamefont {Kroeze},\ and\ \citenamefont {Lukin}}]{lukin2015molecule}%
  \BibitemOpen
  \bibfield  {author} {\bibinfo {author} {\bibfnamefont {Andrei}\ \bibnamefont
  {Derevianko}}, \bibinfo {author} {\bibfnamefont {P{\'e}ter}\ \bibnamefont
  {K{\'o}m{\'a}r}}, \bibinfo {author} {\bibfnamefont {Turker}\ \bibnamefont
  {Topcu}}, \bibinfo {author} {\bibfnamefont {Ronen~M}\ \bibnamefont {Kroeze}},
  \ and\ \bibinfo {author} {\bibfnamefont {Mikhail~D}\ \bibnamefont {Lukin}},\
  }\bibfield  {title} {\enquote {\bibinfo {title} {Effects of molecular
  resonances on rydberg blockade},}\ }\href@noop {} {\bibfield  {journal}
  {\bibinfo  {journal} {Physical Review A}\ }\textbf {\bibinfo {volume} {92}},\
  \bibinfo {pages} {063419} (\bibinfo {year} {2015})}\BibitemShut {NoStop}%
\bibitem [{\citenamefont {Cohen-Tannoudji}\ \emph {et~al.}(1978)\citenamefont
  {Cohen-Tannoudji}, \citenamefont {Hoffbeck},\ and\ \citenamefont
  {Reynaud}}]{CCT1978}%
  \BibitemOpen
  \bibfield  {author} {\bibinfo {author} {\bibfnamefont {C}~\bibnamefont
  {Cohen-Tannoudji}}, \bibinfo {author} {\bibfnamefont {F}~\bibnamefont
  {Hoffbeck}}, \ and\ \bibinfo {author} {\bibfnamefont {S}~\bibnamefont
  {Reynaud}},\ }\bibfield  {title} {\enquote {\bibinfo {title} {Compensating
  {Doppler} broadening with light-shifts},}\ }\href@noop {} {\bibfield
  {journal} {\bibinfo  {journal} {Optics Communications}\ }\textbf {\bibinfo
  {volume} {27}},\ \bibinfo {pages} {71--75} (\bibinfo {year}
  {1978})}\BibitemShut {NoStop}%
\bibitem [{\citenamefont {Yavuz}\ \emph {et~al.}(2013)\citenamefont {Yavuz},
  \citenamefont {Brewer}, \citenamefont {Miles},\ and\ \citenamefont
  {Simmons}}]{yavuz2013suppression}%
  \BibitemOpen
  \bibfield  {author} {\bibinfo {author} {\bibfnamefont {D.~D.}\ \bibnamefont
  {Yavuz}}, \bibinfo {author} {\bibfnamefont {N.~R.}\ \bibnamefont {Brewer}},
  \bibinfo {author} {\bibfnamefont {J.~A.}\ \bibnamefont {Miles}}, \ and\
  \bibinfo {author} {\bibfnamefont {Z.~J.}\ \bibnamefont {Simmons}},\
  }\bibfield  {title} {\enquote {\bibinfo {title} {Suppression of inhomogeneous
  broadening using the ac stark shift},}\ }\href@noop {} {\bibfield  {journal}
  {\bibinfo  {journal} {Physical Review A}\ }\textbf {\bibinfo {volume} {88}},\
  \bibinfo {pages} {063836} (\bibinfo {year} {2013})}\BibitemShut {NoStop}%
\bibitem [{\citenamefont {Reynaud}\ \emph {et~al.}(1979)\citenamefont
  {Reynaud}, \citenamefont {Himbert}, \citenamefont {Dupont-Roc}, \citenamefont
  {Stroke},\ and\ \citenamefont {Cohen-Tannoudji}}]{CCT1979}%
  \BibitemOpen
  \bibfield  {author} {\bibinfo {author} {\bibfnamefont {S}~\bibnamefont
  {Reynaud}}, \bibinfo {author} {\bibfnamefont {M}~\bibnamefont {Himbert}},
  \bibinfo {author} {\bibfnamefont {J}~\bibnamefont {Dupont-Roc}}, \bibinfo
  {author} {\bibfnamefont {HH}~\bibnamefont {Stroke}}, \ and\ \bibinfo {author}
  {\bibfnamefont {C}~\bibnamefont {Cohen-Tannoudji}},\ }\bibfield  {title}
  {\enquote {\bibinfo {title} {Experimental evidence for compensation of
  {Doppler} broadening by light shifts},}\ }\href@noop {} {\bibfield  {journal}
  {\bibinfo  {journal} {Physical Review Letters}\ }\textbf {\bibinfo {volume}
  {42}},\ \bibinfo {pages} {756} (\bibinfo {year} {1979})}\BibitemShut
  {NoStop}%
\bibitem [{\citenamefont {Reynaud}\ \emph {et~al.}(1982)\citenamefont
  {Reynaud}, \citenamefont {Himbert}, \citenamefont {Dalibard}, \citenamefont
  {Dupont-Roc},\ and\ \citenamefont {Cohen-Tannoudji}}]{CCT1982}%
  \BibitemOpen
  \bibfield  {author} {\bibinfo {author} {\bibfnamefont {S}~\bibnamefont
  {Reynaud}}, \bibinfo {author} {\bibfnamefont {M}~\bibnamefont {Himbert}},
  \bibinfo {author} {\bibfnamefont {J}~\bibnamefont {Dalibard}}, \bibinfo
  {author} {\bibfnamefont {J}~\bibnamefont {Dupont-Roc}}, \ and\ \bibinfo
  {author} {\bibfnamefont {C}~\bibnamefont {Cohen-Tannoudji}},\ }\bibfield
  {title} {\enquote {\bibinfo {title} {Compensation of {Doppler} broadening by
  light shifts in two photon absorption},}\ }\href@noop {} {\bibfield
  {journal} {\bibinfo  {journal} {Optics Communications}\ }\textbf {\bibinfo
  {volume} {42}},\ \bibinfo {pages} {39--44} (\bibinfo {year}
  {1982})}\BibitemShut {NoStop}%
\bibitem [{\citenamefont {Kaplan}\ \emph {et~al.}(2002)\citenamefont {Kaplan},
  \citenamefont {Andersen},\ and\ \citenamefont
  {Davidson}}]{kaplan2002suppression}%
  \BibitemOpen
  \bibfield  {author} {\bibinfo {author} {\bibfnamefont {Ariel}\ \bibnamefont
  {Kaplan}}, \bibinfo {author} {\bibfnamefont {Mikkel~Fredslund}\ \bibnamefont
  {Andersen}}, \ and\ \bibinfo {author} {\bibfnamefont {Nir}\ \bibnamefont
  {Davidson}},\ }\bibfield  {title} {\enquote {\bibinfo {title} {Suppression of
  inhomogeneous broadening in rf spectroscopy of optically trapped atoms},}\
  }\href@noop {} {\bibfield  {journal} {\bibinfo  {journal} {Physical Review
  A}\ }\textbf {\bibinfo {volume} {66}},\ \bibinfo {pages} {045401} (\bibinfo
  {year} {2002})}\BibitemShut {NoStop}%
\bibitem [{\citenamefont {Popov}\ and\ \citenamefont
  {Bayev}(2000)}]{popov2000enhanced}%
  \BibitemOpen
  \bibfield  {author} {\bibinfo {author} {\bibfnamefont {Alexander~K}\
  \bibnamefont {Popov}}\ and\ \bibinfo {author} {\bibfnamefont {Alexander~S}\
  \bibnamefont {Bayev}},\ }\bibfield  {title} {\enquote {\bibinfo {title}
  {Enhanced four-wave mixing via elimination of inhomogeneous broadening by
  coherent driving of quantum transitions with control fields},}\ }\href@noop
  {} {\bibfield  {journal} {\bibinfo  {journal} {Physical Review A}\ }\textbf
  {\bibinfo {volume} {62}},\ \bibinfo {pages} {025801} (\bibinfo {year}
  {2000})}\BibitemShut {NoStop}%
\bibitem [{\citenamefont {Arend}\ \emph {et~al.}(2016)\citenamefont {Arend},
  \citenamefont {Becker}, \citenamefont {Sternschulte}, \citenamefont
  {Steinm\"uller-Nethl},\ and\ \citenamefont {Becher}}]{arend2016}%
  \BibitemOpen
  \bibfield  {author} {\bibinfo {author} {\bibfnamefont {Carsten}\ \bibnamefont
  {Arend}}, \bibinfo {author} {\bibfnamefont {Jonas~Nils}\ \bibnamefont
  {Becker}}, \bibinfo {author} {\bibfnamefont {Hadwig}\ \bibnamefont
  {Sternschulte}}, \bibinfo {author} {\bibfnamefont {Doris}\ \bibnamefont
  {Steinm\"uller-Nethl}}, \ and\ \bibinfo {author} {\bibfnamefont {Christoph}\
  \bibnamefont {Becher}},\ }\bibfield  {title} {\enquote {\bibinfo {title}
  {Photoluminescence excitation and spectral hole burning spectroscopy of
  silicon vacancy centers in diamond},}\ }\href {\doibase
  10.1103/PhysRevB.94.045203} {\bibfield  {journal} {\bibinfo  {journal} {Phys.
  Rev. B}\ }\textbf {\bibinfo {volume} {94}},\ \bibinfo {pages} {045203}
  (\bibinfo {year} {2016})}\BibitemShut {NoStop}%
\bibitem [{\citenamefont {Weinzetl}\ \emph {et~al.}(2019)\citenamefont
  {Weinzetl}, \citenamefont {G\"orlitz}, \citenamefont {Becker}, \citenamefont
  {Walmsley}, \citenamefont {Poem}, \citenamefont {Nunn},\ and\ \citenamefont
  {Becher}}]{becher2018coherent}%
  \BibitemOpen
  \bibfield  {author} {\bibinfo {author} {\bibfnamefont {Christian}\
  \bibnamefont {Weinzetl}}, \bibinfo {author} {\bibfnamefont {Johannes}\
  \bibnamefont {G\"orlitz}}, \bibinfo {author} {\bibfnamefont {Jonas~Nils}\
  \bibnamefont {Becker}}, \bibinfo {author} {\bibfnamefont {Ian~A.}\
  \bibnamefont {Walmsley}}, \bibinfo {author} {\bibfnamefont {Eilon}\
  \bibnamefont {Poem}}, \bibinfo {author} {\bibfnamefont {Joshua}\ \bibnamefont
  {Nunn}}, \ and\ \bibinfo {author} {\bibfnamefont {Christoph}\ \bibnamefont
  {Becher}},\ }\bibfield  {title} {\enquote {\bibinfo {title} {Coherent control
  and wave mixing in an ensemble of silicon-vacancy centers in diamond},}\
  }\href {\doibase 10.1103/PhysRevLett.122.063601} {\bibfield  {journal}
  {\bibinfo  {journal} {Phys. Rev. Lett.}\ }\textbf {\bibinfo {volume} {122}},\
  \bibinfo {pages} {063601} (\bibinfo {year} {2019})}\BibitemShut {NoStop}%
\bibitem [{\citenamefont {Happer}(1972)}]{HapperRMP1972}%
  \BibitemOpen
  \bibfield  {author} {\bibinfo {author} {\bibfnamefont {William}\ \bibnamefont
  {Happer}},\ }\bibfield  {title} {\enquote {\bibinfo {title} {Optical
  pumping},}\ }\href {\doibase 10.1103/RevModPhys.44.169} {\bibfield  {journal}
  {\bibinfo  {journal} {Rev. Mod. Phys.}\ }\textbf {\bibinfo {volume} {44}},\
  \bibinfo {pages} {169--249} (\bibinfo {year} {1972})}\BibitemShut {NoStop}%
\bibitem [{\citenamefont {Lezama}\ \emph {et~al.}(1999)\citenamefont {Lezama},
  \citenamefont {Barreiro},\ and\ \citenamefont {Akulshin}}]{Akulshin1999EIA}%
  \BibitemOpen
  \bibfield  {author} {\bibinfo {author} {\bibfnamefont {A.}~\bibnamefont
  {Lezama}}, \bibinfo {author} {\bibfnamefont {S.}~\bibnamefont {Barreiro}}, \
  and\ \bibinfo {author} {\bibfnamefont {A.~M.}\ \bibnamefont {Akulshin}},\
  }\bibfield  {title} {\enquote {\bibinfo {title} {Electromagnetically induced
  absorption},}\ }\href {\doibase 10.1103/PhysRevA.59.4732} {\bibfield
  {journal} {\bibinfo  {journal} {Phys. Rev. A}\ }\textbf {\bibinfo {volume}
  {59}},\ \bibinfo {pages} {4732--4735} (\bibinfo {year} {1999})}\BibitemShut
  {NoStop}%
\bibitem [{\citenamefont {Taichenachev}\ \emph {et~al.}(1999)\citenamefont
  {Taichenachev}, \citenamefont {Tumaikin},\ and\ \citenamefont
  {Yudin}}]{Yudin1999TOC}%
  \BibitemOpen
  \bibfield  {author} {\bibinfo {author} {\bibfnamefont {A.~V.}\ \bibnamefont
  {Taichenachev}}, \bibinfo {author} {\bibfnamefont {A.~M.}\ \bibnamefont
  {Tumaikin}}, \ and\ \bibinfo {author} {\bibfnamefont {V.~I.}\ \bibnamefont
  {Yudin}},\ }\bibfield  {title} {\enquote {\bibinfo {title}
  {Electromagnetically induced absorption in a four-state system},}\ }\href
  {\doibase 10.1103/PhysRevA.61.011802} {\bibfield  {journal} {\bibinfo
  {journal} {Phys. Rev. A}\ }\textbf {\bibinfo {volume} {61}},\ \bibinfo
  {pages} {011802} (\bibinfo {year} {1999})}\BibitemShut {NoStop}%
\bibitem [{\citenamefont {Goren}\ \emph {et~al.}(2003)\citenamefont {Goren},
  \citenamefont {Wilson-Gordon}, \citenamefont {Rosenbluh},\ and\ \citenamefont
  {Friedmann}}]{Goren2003EIA}%
  \BibitemOpen
  \bibfield  {author} {\bibinfo {author} {\bibfnamefont {C.}~\bibnamefont
  {Goren}}, \bibinfo {author} {\bibfnamefont {A.~D.}\ \bibnamefont
  {Wilson-Gordon}}, \bibinfo {author} {\bibfnamefont {M.}~\bibnamefont
  {Rosenbluh}}, \ and\ \bibinfo {author} {\bibfnamefont {H.}~\bibnamefont
  {Friedmann}},\ }\bibfield  {title} {\enquote {\bibinfo {title}
  {Electromagnetically induced absorption due to transfer of coherence and to
  transfer of population},}\ }\href {\doibase 10.1103/PhysRevA.67.033807}
  {\bibfield  {journal} {\bibinfo  {journal} {Phys. Rev. A}\ }\textbf {\bibinfo
  {volume} {67}},\ \bibinfo {pages} {033807} (\bibinfo {year}
  {2003})}\BibitemShut {NoStop}%
\bibitem [{\citenamefont {Tilchin}\ \emph {et~al.}(2011)\citenamefont
  {Tilchin}, \citenamefont {Wilson-Gordon},\ and\ \citenamefont
  {Firstenberg}}]{TilchinPRA2011}%
  \BibitemOpen
  \bibfield  {author} {\bibinfo {author} {\bibfnamefont {E.}~\bibnamefont
  {Tilchin}}, \bibinfo {author} {\bibfnamefont {A.~D.}\ \bibnamefont
  {Wilson-Gordon}}, \ and\ \bibinfo {author} {\bibfnamefont {O.}~\bibnamefont
  {Firstenberg}},\ }\bibfield  {title} {\enquote {\bibinfo {title} {Effects of
  thermal motion on electromagnetically induced absorption},}\ }\href {\doibase
  10.1103/PhysRevA.83.053812} {\bibfield  {journal} {\bibinfo  {journal} {Phys.
  Rev. A}\ }\textbf {\bibinfo {volume} {83}},\ \bibinfo {pages} {053812}
  (\bibinfo {year} {2011})}\BibitemShut {NoStop}%
\bibitem [{\citenamefont {Bae}\ \emph {et~al.}(2010)\citenamefont {Bae},
  \citenamefont {Moon}, \citenamefont {Kim}, \citenamefont {Lee},\ and\
  \citenamefont {Kim}}]{Kim2010EIA}%
  \BibitemOpen
  \bibfield  {author} {\bibinfo {author} {\bibfnamefont {In-Ho}\ \bibnamefont
  {Bae}}, \bibinfo {author} {\bibfnamefont {Han~Seb}\ \bibnamefont {Moon}},
  \bibinfo {author} {\bibfnamefont {Min-Koeung}\ \bibnamefont {Kim}}, \bibinfo
  {author} {\bibfnamefont {Lim}\ \bibnamefont {Lee}}, \ and\ \bibinfo {author}
  {\bibfnamefont {Jung~Bog}\ \bibnamefont {Kim}},\ }\bibfield  {title}
  {\enquote {\bibinfo {title} {Transformation of electromagnetically induced
  transparency into enhanced absorption with a standing-wave coupling field in
  an {Rb} vapor cell},}\ }\href {\doibase 10.1364/OE.18.001389} {\bibfield
  {journal} {\bibinfo  {journal} {Opt. Express}\ }\textbf {\bibinfo {volume}
  {18}},\ \bibinfo {pages} {1389--1397} (\bibinfo {year} {2010})}\BibitemShut
  {NoStop}%
\bibitem [{\citenamefont {Bason}\ \emph {et~al.}(2009)\citenamefont {Bason},
  \citenamefont {Mohapatra}, \citenamefont {Weatherill},\ and\ \citenamefont
  {Adams}}]{Adams2009Nsystem}%
  \BibitemOpen
  \bibfield  {author} {\bibinfo {author} {\bibfnamefont {M~G}\ \bibnamefont
  {Bason}}, \bibinfo {author} {\bibfnamefont {A~K}\ \bibnamefont {Mohapatra}},
  \bibinfo {author} {\bibfnamefont {K~J}\ \bibnamefont {Weatherill}}, \ and\
  \bibinfo {author} {\bibfnamefont {C~S}\ \bibnamefont {Adams}},\ }\bibfield
  {title} {\enquote {\bibinfo {title} {Narrow absorptive resonances in a
  four-level atomic system},}\ }\href {\doibase 10.1088/0953-4075/42/7/075503}
  {\bibfield  {journal} {\bibinfo  {journal} {Journal of Physics B: Atomic,
  Molecular and Optical Physics}\ }\textbf {\bibinfo {volume} {42}},\ \bibinfo
  {pages} {075503} (\bibinfo {year} {2009})}\BibitemShut {NoStop}%
\bibitem [{\citenamefont {Whiting}\ \emph {et~al.}(2015)\citenamefont
  {Whiting}, \citenamefont {Bimbard}, \citenamefont {Keaveney}, \citenamefont
  {Zentile}, \citenamefont {Adams},\ and\ \citenamefont
  {Hughes}}]{Whiting2015EIA}%
  \BibitemOpen
  \bibfield  {author} {\bibinfo {author} {\bibfnamefont {Daniel~J.}\
  \bibnamefont {Whiting}}, \bibinfo {author} {\bibfnamefont {Erwan}\
  \bibnamefont {Bimbard}}, \bibinfo {author} {\bibfnamefont {James}\
  \bibnamefont {Keaveney}}, \bibinfo {author} {\bibfnamefont {Mark~A.}\
  \bibnamefont {Zentile}}, \bibinfo {author} {\bibfnamefont {Charles~S.}\
  \bibnamefont {Adams}}, \ and\ \bibinfo {author} {\bibfnamefont {Ifan~G.}\
  \bibnamefont {Hughes}},\ }\bibfield  {title} {\enquote {\bibinfo {title}
  {Electromagnetically induced absorption in a nondegenerate three-level ladder
  system},}\ }\href {\doibase 10.1364/OL.40.004289} {\bibfield  {journal}
  {\bibinfo  {journal} {Opt. Lett.}\ }\textbf {\bibinfo {volume} {40}},\
  \bibinfo {pages} {4289--4292} (\bibinfo {year} {2015})}\BibitemShut {NoStop}%
\bibitem [{\citenamefont {Finkelstein}\ \emph {et~al.}(2019)\citenamefont
  {Finkelstein}, \citenamefont {Lahad}, \citenamefont {Michel}, \citenamefont
  {Davidson}, \citenamefont {Poem},\ and\ \citenamefont
  {Firstenberg}}]{finkelstein2019twocolor}%
  \BibitemOpen
  \bibfield  {author} {\bibinfo {author} {\bibfnamefont {Ran}\ \bibnamefont
  {Finkelstein}}, \bibinfo {author} {\bibfnamefont {Ohr}\ \bibnamefont
  {Lahad}}, \bibinfo {author} {\bibfnamefont {Ohad}\ \bibnamefont {Michel}},
  \bibinfo {author} {\bibfnamefont {Omri}\ \bibnamefont {Davidson}}, \bibinfo
  {author} {\bibfnamefont {Eilon}\ \bibnamefont {Poem}}, \ and\ \bibinfo
  {author} {\bibfnamefont {Ofer}\ \bibnamefont {Firstenberg}},\ }\bibfield
  {title} {\enquote {\bibinfo {title} {Power narrowing: counteracting doppler
  broadening in two-color transitions},}\ }\href@noop {} {\bibfield  {journal}
  {\bibinfo  {journal} {New Journal of Physics}\ }\textbf {\bibinfo {volume}
  {21}},\ \bibinfo {pages} {103024} (\bibinfo {year} {2019})}\BibitemShut
  {NoStop}%
\bibitem [{\citenamefont {Wade}\ \emph {et~al.}(2018)\citenamefont {Wade},
  \citenamefont {Marcuzzi}, \citenamefont {Levi}, \citenamefont {Kondo},
  \citenamefont {Lesanovsky}, \citenamefont {Adams},\ and\ \citenamefont
  {Weatherill}}]{wade2018terahertz}%
  \BibitemOpen
  \bibfield  {author} {\bibinfo {author} {\bibfnamefont {Christopher~G}\
  \bibnamefont {Wade}}, \bibinfo {author} {\bibfnamefont {Matteo}\ \bibnamefont
  {Marcuzzi}}, \bibinfo {author} {\bibfnamefont {Emanuele}\ \bibnamefont
  {Levi}}, \bibinfo {author} {\bibfnamefont {Jorge~M}\ \bibnamefont {Kondo}},
  \bibinfo {author} {\bibfnamefont {Igor}\ \bibnamefont {Lesanovsky}}, \bibinfo
  {author} {\bibfnamefont {Charles~S}\ \bibnamefont {Adams}}, \ and\ \bibinfo
  {author} {\bibfnamefont {Kevin~J}\ \bibnamefont {Weatherill}},\ }\bibfield
  {title} {\enquote {\bibinfo {title} {A terahertz-driven non-equilibrium phase
  transition in a room temperature atomic vapour},}\ }\href@noop {} {\bibfield
  {journal} {\bibinfo  {journal} {Nature communications}\ }\textbf {\bibinfo
  {volume} {9}},\ \bibinfo {pages} {3567} (\bibinfo {year} {2018})}\BibitemShut
  {NoStop}%
\bibitem [{\citenamefont {Kara}\ \emph {et~al.}(2018)\citenamefont {Kara},
  \citenamefont {Bhowmick},\ and\ \citenamefont
  {Mohapatra}}]{Mohapatra2018rydberg}%
  \BibitemOpen
  \bibfield  {author} {\bibinfo {author} {\bibfnamefont {Dushmanta}\
  \bibnamefont {Kara}}, \bibinfo {author} {\bibfnamefont {Arup}\ \bibnamefont
  {Bhowmick}}, \ and\ \bibinfo {author} {\bibfnamefont {Ashok~K}\ \bibnamefont
  {Mohapatra}},\ }\bibfield  {title} {\enquote {\bibinfo {title} {Rydberg
  interaction induced enhanced excitation in thermal atomic vapor},}\
  }\href@noop {} {\bibfield  {journal} {\bibinfo  {journal} {Scientific
  reports}\ }\textbf {\bibinfo {volume} {8}},\ \bibinfo {pages} {5256}
  (\bibinfo {year} {2018})}\BibitemShut {NoStop}%
\bibitem [{\citenamefont {Levine}\ \emph {et~al.}(2018)\citenamefont {Levine},
  \citenamefont {Keesling}, \citenamefont {Omran}, \citenamefont {Bernien},
  \citenamefont {Schwartz}, \citenamefont {Zibrov}, \citenamefont {Endres},
  \citenamefont {Greiner}, \citenamefont {Vuleti\ifmmode~\acute{c}\else
  \'{c}\fi{}},\ and\ \citenamefont {Lukin}}]{Lukin2018fidelity}%
  \BibitemOpen
  \bibfield  {author} {\bibinfo {author} {\bibfnamefont {Harry}\ \bibnamefont
  {Levine}}, \bibinfo {author} {\bibfnamefont {Alexander}\ \bibnamefont
  {Keesling}}, \bibinfo {author} {\bibfnamefont {Ahmed}\ \bibnamefont {Omran}},
  \bibinfo {author} {\bibfnamefont {Hannes}\ \bibnamefont {Bernien}}, \bibinfo
  {author} {\bibfnamefont {Sylvain}\ \bibnamefont {Schwartz}}, \bibinfo
  {author} {\bibfnamefont {Alexander~S.}\ \bibnamefont {Zibrov}}, \bibinfo
  {author} {\bibfnamefont {Manuel}\ \bibnamefont {Endres}}, \bibinfo {author}
  {\bibfnamefont {Markus}\ \bibnamefont {Greiner}}, \bibinfo {author}
  {\bibfnamefont {Vladan}\ \bibnamefont {Vuleti\ifmmode~\acute{c}\else
  \'{c}\fi{}}}, \ and\ \bibinfo {author} {\bibfnamefont {Mikhail~D.}\
  \bibnamefont {Lukin}},\ }\bibfield  {title} {\enquote {\bibinfo {title}
  {High-fidelity control and entanglement of rydberg-atom qubits},}\ }\href
  {\doibase 10.1103/PhysRevLett.121.123603} {\bibfield  {journal} {\bibinfo
  {journal} {Phys. Rev. Lett.}\ }\textbf {\bibinfo {volume} {121}},\ \bibinfo
  {pages} {123603} (\bibinfo {year} {2018})}\BibitemShut {NoStop}%
\bibitem [{\citenamefont {Murray}\ \emph {et~al.}(2018)\citenamefont {Murray},
  \citenamefont {Mirgorodskiy}, \citenamefont {Tresp}, \citenamefont {Braun},
  \citenamefont {Paris-Mandoki}, \citenamefont {Gorshkov}, \citenamefont
  {Hofferberth},\ and\ \citenamefont {Pohl}}]{hofferberth2018subtraction}%
  \BibitemOpen
  \bibfield  {author} {\bibinfo {author} {\bibfnamefont {C.~R.}\ \bibnamefont
  {Murray}}, \bibinfo {author} {\bibfnamefont {I.}~\bibnamefont
  {Mirgorodskiy}}, \bibinfo {author} {\bibfnamefont {C.}~\bibnamefont {Tresp}},
  \bibinfo {author} {\bibfnamefont {C.}~\bibnamefont {Braun}}, \bibinfo
  {author} {\bibfnamefont {A.}~\bibnamefont {Paris-Mandoki}}, \bibinfo {author}
  {\bibfnamefont {A.~V.}\ \bibnamefont {Gorshkov}}, \bibinfo {author}
  {\bibfnamefont {S.}~\bibnamefont {Hofferberth}}, \ and\ \bibinfo {author}
  {\bibfnamefont {T.}~\bibnamefont {Pohl}},\ }\bibfield  {title} {\enquote
  {\bibinfo {title} {Photon subtraction by many-body decoherence},}\ }\href
  {\doibase 10.1103/PhysRevLett.120.113601} {\bibfield  {journal} {\bibinfo
  {journal} {Phys. Rev. Lett.}\ }\textbf {\bibinfo {volume} {120}},\ \bibinfo
  {pages} {113601} (\bibinfo {year} {2018})}\BibitemShut {NoStop}%
\bibitem [{\citenamefont {Busche}\ \emph {et~al.}(2017)\citenamefont {Busche},
  \citenamefont {Huillery}, \citenamefont {Ball}, \citenamefont {Ilieva},
  \citenamefont {Jones},\ and\ \citenamefont {Adams}}]{adams2017contactless}%
  \BibitemOpen
  \bibfield  {author} {\bibinfo {author} {\bibfnamefont {Hannes}\ \bibnamefont
  {Busche}}, \bibinfo {author} {\bibfnamefont {Paul}\ \bibnamefont {Huillery}},
  \bibinfo {author} {\bibfnamefont {Simon~W}\ \bibnamefont {Ball}}, \bibinfo
  {author} {\bibfnamefont {Teodora}\ \bibnamefont {Ilieva}}, \bibinfo {author}
  {\bibfnamefont {Matthew~PA}\ \bibnamefont {Jones}}, \ and\ \bibinfo {author}
  {\bibfnamefont {Charles~S}\ \bibnamefont {Adams}},\ }\bibfield  {title}
  {\enquote {\bibinfo {title} {Contactless nonlinear optics mediated by
  long-range rydberg interactions},}\ }\href@noop {} {\bibfield  {journal}
  {\bibinfo  {journal} {Nature Physics}\ }\textbf {\bibinfo {volume} {13}},\
  \bibinfo {pages} {655} (\bibinfo {year} {2017})}\BibitemShut {NoStop}%
\end{thebibliography}%

\end{document}